\documentclass[fleqn,usenatbib]{mnras}

\usepackage{newtxtext,newtxmath}

\usepackage[T1]{fontenc}
\usepackage{ae,aecompl}

\usepackage{graphicx}	
\usepackage{amsmath}	
\usepackage{amssymb}	
\usepackage{xspace}
\usepackage{amsfonts,textcomp}
\usepackage[usenames]{color}
\usepackage{times}
\usepackage{hyperref}
\usepackage{ulem}
\usepackage{enumitem}
\usepackage[compatibility=false]{caption}
\usepackage{subcaption}
\usepackage{multirow}
\usepackage{tabulary}
\usepackage[para]{threeparttable}
\usepackage{array,booktabs,longtable,tabularx}
\newcolumntype{L}{>{\raggedright\arraybackslash}X}
\usepackage{ltablex}

\usepackage{etoolbox}
\makeatletter
\patchcmd\@combinedblfloats{\box\@outputbox}{\unvbox\@outputbox}{}{%
   \errmessage{\noexpand\@combinedblfloats could not be patched}%
}%
 \makeatother

\renewlist{tablenotes}{enumerate}{1}
\makeatletter
\setlist[tablenotes]{label=\tnote{\alph*},ref=\alph*,itemsep=\z@,topsep=\z@skip,partopsep=\z@skip,parsep=\z@,itemindent=\z@,labelindent=\tabcolsep,labelsep=.2em,leftmargin=*,align=left,before={\footnotesize}}
\makeatother

\newcommand{\msun}{\mbox{$\rm{M}_\odot$}\xspace}
\newcommand{\disperse}{\mbox{{\sc DisPerSE}}\xspace}
\newcommand{\ssfr}{\mbox{sSFR}\xspace}
\newcommand{\mhalo}{\mbox{$M_h$}\xspace}
\newcommand{\vsig}{\mbox{$v/\sigma$}\xspace}
\newcommand{\mstar}{\mbox{$M_\star$}\xspace}
\newcommand{\hagn}{\mbox{{\sc Horizon-AGN}}\xspace}
\newcommand{\hnoagn}{\mbox{{\sc  Horizon-noAGN}}\xspace}
\newcommand{\simba}{\mbox{{\sc Simba}}\xspace}
\newcommand{\mufasa}{\mbox{{\sc Mufasa}}\xspace}
\newcommand{\gizmo}{\mbox{{\sc  Gizmo}}\xspace}
\newcommand{\ramses}{\mbox{{\sc  Ramses}}\xspace}
\newcommand{\yt}{\mbox{{\sc YT}}\xspace}
\newcommand{\caesar}{\mbox{{\sc Caesar}}\xspace}

\definecolor{Orange}{rgb}{1.0,0.5,0.15}
\definecolor{Blue}{rgb}{0,0.08,0.65}
\definecolor{Red}{rgb}{0.65,0.08,0.05}
\definecolor{Green}{rgb}{0.15,0.45,0.25}
\definecolor{Pink}{rgb}{1.0,0.05,0.5}

\title[The impact of connectivity on galaxy properties]{The impact of the connectivity of the cosmic web\\
on the physical properties of galaxies at its nodes}

\author[K. Kraljic, C. Pichon, S. Codis et al.
]{\parbox[t]{\textwidth}{
Katarina Kraljic$^{1}$\thanks{E-mail: kat@roe.ac.uk}, Christophe Pichon$^{2,3}$, Sandrine Codis$^{2}$, Clotilde Laigle$^{2}$, \\
Romeel Dav\'e$^{1,4,5}$,  
Yohan Dubois$^{2}$, Ho Seong Hwang$^{6}$, Dmitri Pogosyan$^{7}$, \\
St\'ephane Arnouts$^{8}$,
Julien Devriendt$^{9}$, Marcello Musso$^{10}$, 
S\'ebastien Peirani$^{2,11}$, \\
Adrianne Slyz$^{9}$, Marie Treyer$^{8}$
}
\\
\\
$^{1}$Institute for Astronomy, University of Edinburgh, Royal Observatory, Blackford Hill, Edinburgh, EH9 3HJ, United Kingdom\\ 
$^{2}$ CNRS and Sorbonne Universit\'e, UMR 7095, Institut d'Astrophysique de Paris, 98 bis Boulevard Arago, F-75014 Paris, France\\
$^{3}$ School of Physics, Korea Institute for Advanced Study (KIAS), 85 Hoegiro, Dongdaemun-gu, Seoul, 02455, Republic of Korea\\
$^{4}$ University of the Western Cape, Bellville, Cape Town 7535, South Africa\\
$^{5}$ South African Astronomical Observatories, Observatory, Cape Town 7925, South Africa\\
$^{6}$ Korea Astronomy and Space Science Institute (KASI), 776 Daedeokdae-ro, Yuseong-gu, Daejeon 34055, Republic of Korea\\
$^{7}$ Department of Physics, University of Alberta, 412 Avadh Bhatia Physics Laboratory, Edmonton, Alberta, T6G 2J1, Canada\\
$^{8}$ Aix Marseille Universit\'e, CNRS, Laboratoire d'Astrophysique de Marseille, UMR 7326, F-13388, Marseille, France\\
$^{9}$ Department of Physics, University of Oxford, Keble Road, Oxford, OX1 3RH,United Kingdom\\
$^{10}$ East African Institute for Fundamental Research (ICTP-EAIFR), KIST2 Building, Nyarugenge Campus, University of Rwanda, Kigali, Rwanda\\
$^{11}$ Observatoire de la C\^ote d'Azur, CNRS, Laboratoire Lagrange, Bd de l'Observatoire, CS 34229, F-06304 Nice Cedex 4, France
}

\date{Accepted XXX. Received YYY; in original form ZZZ}

\pubyear{2019}

\begin{document}
\label{firstpage}
\pagerange{\pageref{firstpage}--\pageref{lastpage}}
\maketitle

\begin{abstract}
We investigate the impact of the number of filaments connected to the nodes of the cosmic web on the physical properties of their galaxies using the Sloan Digital Sky Survey. We compare these measurements to  the cosmological hydrodynamical simulations Horizon-(no)AGN and \simba. We find that more massive galaxies are more connected, in qualitative agreement with theoretical predictions and measurements in dark matter only simulation.
The star formation activity and morphology of observed galaxies both display some dependence on the connectivity of the cosmic web at fixed stellar mass: less star forming and less rotation supported galaxies also tend to have higher connectivity. 
These results qualitatively hold both for observed and virtual galaxies, and can be understood given that  the cosmic web is the main source of fuel for galaxy growth. 
The simulations show the same trends at fixed halo mass, suggesting that the geometry of filamentary infall impacts galaxy properties beyond the depth of the local potential well. Based on simulations, it is also found that AGN feedback is key in reversing the relationship between stellar mass and connectivity at fixed halo mass.
Technically, connectivity is a practical observational proxy for past and present accretion (minor mergers or diffuse infall).
\end{abstract}



\section{Introduction}

During the past few decades, the  $\Lambda$CDM  concordant  model  has been established as a framework of choice in which to interpret how and when  galaxies acquire their physical properties. The cosmological model predicts a certain shape for the initial power spectrum of density fluctuations, leading to the hierarchical formation of the large-scale structure, as observed more than 30 years ago by the first CfA catalogue \citep{Lapparent1986}. This so-called 'cosmic web' \citep{Klypin1993,bondetal1996} connects the observed clusters of galaxies via a filamentary network arising from the geometrical properties of the initial density field  enhanced by anisotropic gravitational collapse \citep{lynden-bell64,zeldovich70}. 
Arguably one of the most important features of this framework is to also provide an explanation of the correlation of many galactic properties (kinematics, star formation rates) beyond galactic mass.
The favoured culprit is unsurprisingly the interplay between galaxies and the geometry and content of the intergalactic medium at large. The large-scale environment of galaxies seems to play a significant role in shaping some of their properties e.g. through torques, while the rest of them are thought to depend mostly on small-scale (internal) processes. 

Since the mid 1970s there have been many studies devoted to measuring the impact of environment on galaxy properties. \cite{davis1976} first pointed out that early-type galaxies are more strongly clustered than  late types, while \cite{Dressler1980} identified a morphology-density relation: galaxies living in denser environments tend to be redder and have lower star formation rates (SFRs) than their  isolated counterparts. 
Since then, a variety of density relations, such as color-density, star formation-density, morphology-density relation, has been pointed out  by many other works 
\citep[e.g.][and references therein]{Dressler1997,Hashimoto1998,Lewis2002,Goto2003,Blanton2003,Baldry2006,Bamford2009,Cucciati2010,Burton2013,Cucciati2017}.

\begin{figure*}
\centering\includegraphics[width=2\columnwidth]{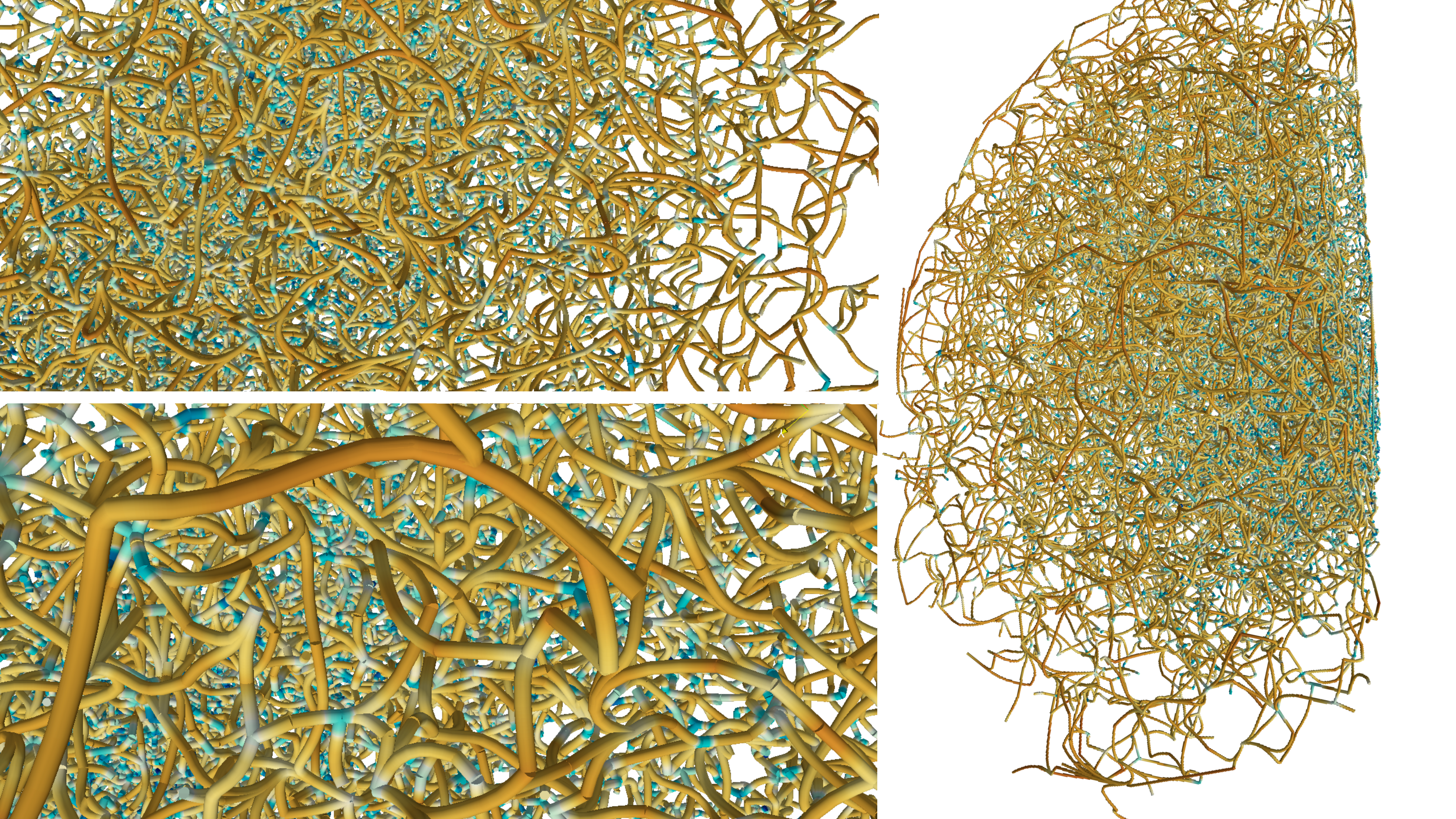}
\caption{The skeleton of the SDSS DR7 traced by \disperse with a persistence level of 3$\sigma$ from the full galaxy sample at $z \leq 0.19$. 
{\sl Left-hand panel}: zoomed-in regions.
{\sl Right-hand panel}: the region over which the connectivity of the network was computed.
The colour code traces the underlying density, while the width represents the robustness of the filament.
This \protect\href{http://3dviewer.horizon-simulation.org/3dclouds/SDSS_DR10_segs_graph_points.html}{url}
provides us with a 3D rendering of the SDSS skeleton.
}
\label{fig:tempel-skeleton}
\end{figure*} 

A fundamental requirement to understand such galaxy properties beyond mass is to properly characterize the geometry of their environment,  spanning as broad a range as possible, from  field galaxies to groups and rich clusters. The vast majority of the studies in the literature accomplishes this task either by counting the number of neighbours that a galaxy has within a fixed aperture on the sky, or by measuring the distance to the n$^{th}$ nearest neighbour. Although these indicators are easy to obtain, their physical interpretation is far from straightforward \citep[e.g.][]{Blanton2003,Kauffmann2004}. 

A novel approach was introduced by \cite{codisetal2018}, who focused on the connectivity of the cosmic web network as a means to understand its morphology and geometry. The motivation was two-fold: i) to characterize the underlying cosmology, since the disconnection of filaments is driven by gravitational clustering and by dark energy which will stretch and disconnect neighbouring filaments; and ii) to probe the geometry of accretion on halo and galactic scales. In particular, they showed that connectivity is an increasing function of halo mass, a property that is directly inherited from the underlying density peaks whose connectivity increases with their height\footnote{The higher the peaks the larger the connectivity because high peaks are more isotropic.}.
On smaller scales, cosmic connectivity provides a unique view into
galactic assembly history, since dark matter filaments typically have a baryonic continuation within dark matter haloes. This closely connects the cosmic environment to the growth of embedded galaxies \citep[e.g.][]{Keres2005,Dekel2009,Keres2009}.
In this context, \cite{DarraghFord2019} measured the local connectivity, called the multiplicity of connected filaments \citep{codisetal2018}, 
in the  COSMOS field around X-ray detected groups at higher redshifts ($0.5 < z < 1.2$). 
To do so, they extracted cosmic filaments using the photometric redshifts of the COSMOS2015 catalogue \citep{Laigle2016} in two-dimensional slices \citep[see also][]{Sarron2019}.

The aim of the current project is to measure the connectivity, both global and local, of the cosmic web in 3D as traced by galaxies in the SDSS survey \citep{SDSS2000,SDSS}, 
and to study how it relates to the physical properties of galaxies identified as nodes. 
In particular, we assess the impact of connectivity on the star formation activity of galaxies and the role it may play in quenching star formation. Using  a set of cosmological simulations, our aim is thus to address the question: at fixed dark halo mass, to what extent is galactic connectivity a
driver of stellar mass, star formation activity, kinematics or spiral fraction? 

The outline of the paper is the following.
Section~\ref{sec:data} presents the simulated and observed data sets used in this work.
Section~\ref{sec:results} presents the local and global connectivity statistics, while
Section~\ref{sec:predictions} compares our findings to 
the connectivity of simulated galaxies.
Section~\ref{sec:discussion} investigates the evolution of connectivity and the physical properties of galaxies at fixed halo mass. 
Finally, Section~\ref{sec:conclusions} wraps up.
Appendix~\ref{sec:persistence} presents the impact of persistence on connectivity, while appendix~\ref{sec:satellite-mass} focuses on the effect of satellites on connectivity. Appendix~\ref{sec:multiplicity} contains complementary measurements of the multiplicity. Finally,  Appendix~\ref{sec:simba_results} contains complementary measurements in the \simba simulation. 

Throughout this paper, by log, we refer to the 10-based logarithm. If not stated differently, statistical errors are computed by bootstrapping, such that the errors on a given statistical quantity correspond to the standard deviation of the distribution of that quantity re-computed in 100 random samples drawn from the parent sample with replacement.

\section{Data and methods}
\label{sec:data} 
Let us first describe the data sets and 
the methods used to characterise the connectivity of observed and virtual surveys.
Section~\ref{sec:obs-data} presents the SDSS catalogue used to compute the filaments,
Section~\ref{sec:cw} describes the algorithm used to extract them, 
while Section~\ref{sec:simus} presents the three hydrodynamical simulations used in the comparison with the observations.

\subsection{Observational data}
\label{sec:obs-data}

\subsubsection{SDSS catalogues}

We use spectroscopic data from the main galaxy sample of the SDSS data release 7 (DR7, \citealt{sdssdr7}). The main galaxy sample includes photometric and spectroscopic data for $\sim 7 \times10^5$ galaxies at $m_r < 17.77$. However, spectroscopic completeness for bright galaxies (e.g. $m_r < 14.5$) is poor because of saturation and cross-talk in the spectrograph, and also for the galaxies in high-density regions including galaxy clusters because of fiber collisions. 
Therefore, \citet{choi10} compiled redshift information from the literature for the galaxies at $m_r < 17.77$ that were missed in the original SDSS galaxy catalogue. We use this Korea Institute for Advanced Study (KIAS) DR7 value-added galaxy catalogue \href{http://astro.kias.re.kr/vagc/dr7/}{KIAS-VAGC} for further analysis.

To reduce the effect of Fingers-of-God along the line-of-sight introduced by peculiar motions in galaxy groups or clusters on the identification of large-scale structures, we use a method similar to the ones adopted in \citet[see also \citealt{park12lss,hwang16}]{teg04}. 
We first run the Friends-of-Friends algorithm with a linking length of 3 $h^{-1}$ Mpc (roughly corresponding to the size of a galaxy cluster) to identify group-like structures.
We then compare the dispersion of the identified structures perpendicular and parallel to the line-of-sight. 
If the dispersion parallel to the line-of-sight is larger than the perpendicular one, we revise the radial velocities of the galaxies within the structures to have the same velocity dispersion in the two directions.

We adopt the SFRs and stellar masses from the \href{http://www.mpa-garching.mpg.de/SDSS/DR7/sfrs.html}{MPA/JHU DR7 VAGC}.
The SFRs are extinction and aperture corrected ones derived from the SDSS spectra.
For star-forming galaxies classified in the emission line-ratio diagram, they use the H$\alpha$ emission line luminosity to derive SFR. For all other galaxies including AGN-host galaxies where the SFRs cannot be directly measured, the 4000\AA~break (D4000) is used for SFR estimates (see \citealt{bri04} for details).
The stellar masses are derived from the fit to the SDSS five-band photometric data with the model of \citet[see also \citealt{kau03}]{bc03}.
The SFR and stellar mass estimates in the MPA/JHU DR7 VAGC are based on the Kroupa initial mass function (IMF, \citealt{kro01}).

The galaxy morphology information is taken from the KIAS DR7 VAGC \citep{choi10}. In this catalogue, galaxies are initially classified into two types based on ($u-r$) colour, ($g-i$) colour gradient and $i$-band concentration index \citep{pc05}: early (ellipticals and lenticulars) and late (spirals and irregulars) types. The resulting completeness and reliability for the morphological classification reaches 90 \%.
To complement this automated classification scheme, 13 astronomers in the KIAS group performed an additional visual check of the SDSS $gri$ colour images for the galaxies misclassified by the automated scheme.
During this inspection, they revised the morphological types of blended or merging galaxies, blue but elliptical-shaped galaxies, and dusty edge-on spirals.
It was found that the galaxy morphology in the KIAS VAGC agrees well with that in the Galaxy Zoo catalogues \citep[][]{Lin11,Wil13} for $\sim$ 81 \% of the SDSS main galaxy sample.

The classification between the star-forming and passive populations is based on a \ssfr and \ssfr-\mstar cut. Galaxies are classified as passive if their \ssfr $\leq 10^{-11}$ yr$^{-1}$ or  \ssfr $\leq 10^{-10.6}$ yr$^{-1}$ and $\log(\ssfr$/yr$^{-1}$) $\leq -10.7 - 0.2 \times [\log(\mstar/\msun) - 9.5]$ \citep[see][]{Moustakas2013}.

\subsection{The cosmic web of the SDSS}
\label{sec:cw}

In this paper, we measure the connectivity of the cosmic web
using the 3D ridge extractor \disperse \citep{sousbie112}\footnote{This paper provided a first analysis of all topological components of the SDSS' cosmic web, while \cite{Gay2010} first analysed galactic properties as a function of the distance to filaments and nodes in a cosmological hydrodynamical simulation.}, 
which identifies the so-called skeleton (critical lines connecting peaks together) as 
1D-ascending manifolds of the discrete Morse-Smale complex \citep{Forman2002}. 
This software has been shown to provide a consistent estimate 
for the connectivity of dark matter peaks \citep{codisetal2018}, and to operate well on discrete inhomogeneous galaxy catalogues \citep[e.g.][]{Malavasi2017,Kraljic2018,Laigle2018}.
This scale-free algorithm relies on topological persistence 
as a mean to identify robust\footnote{Here robust refers to the \href{https://en.wikipedia.org/wiki/Topology}{topological} nature of ridges,
see Section~\ref{sec:discussion}.} components of the 
filamentary network, quantified in terms of significance 
compared to a discrete random Poisson distribution, through the so-called persistence $N_\sigma$\footnote{Increasing the persistence threshold allows to eliminate less significant critical pairs and to retain only the most topologically robust features.}.
In this paper, all results are shown for $N_\sigma=3$, unless stated otherwise. 
We have checked that the choice of the threshold does not alter our conclusions (see Appendix~\ref{sec:persistence}). 
Figure~\ref{fig:tempel-skeleton} displays the corresponding skeleton constructed from the galaxy distribution. 

In \disperse, the connectivity is computed automatically: all ridges connecting one peak to its saddles are identified  by construction, since the algorithm
simplifies specifically components of the tessellation connecting peaks to saddle points using persistence ratios. The number of connected ridges is in practice
stored at each node.  The multiplicity (or local connectivity)
is also straightforwardly extracted from  the skeleton via the identification of the bifurcation points of the skeleton \citep{Pogosyan2009}. The multiplicity, $\mu$ is then simply the connectivity, $\kappa$, minus the number of bifurcation points associated with each node. 
Connectivity and multiplicity are complementary since they probe different scales: connectivity is a measure of the larger-scale topology of the environment while multiplicity focuses on the number of {\it local} connected filaments, regardless of whether or not those filaments will bifurcate and split in two or more further away from the central node. 
As such, multiplicity is more closely related to the local accretion of matter.

In this work, we consider the properties of galaxies that have been identified by \disperse as nodes of the cosmic web, and investigate how they relate to connectivity (global and local)\footnote{In practice, each node of the cosmic web is associated with its closest galaxy. Ideally one should distinguish between centrals and satellites, however observationally this is challenging. Since the nodes of the cosmic web are the peaks of the density field, at high mass, picking the closest galaxy often yields the central galaxy. At low mass however the bright central galaxy is not always sitting at the center of mass of the group and therefore the closest galaxy from the node can be a satellite. This impacts the interpretation mostly at the low-mass end, see Appendix~\ref{sec:satellite-mass}.}.

\subsection{Cosmological simulations}
\label{sec:simus}

In order to make robust predictions for the impact of connectivity on galaxy properties, we examine two independent cosmological galaxy formation simulations, one based on adaptive mesh refinement and the other using a meshless hydrodynamics method. The analysis of these simulations is performed at redshift $z=0$.

\subsubsection{The Horizon-AGN AMR cosmological simulation}

The \hagn\footnote{\href{http://www.horizon-simulation.org/}{www.horizon-simulation.org}} simulation is described in detail in 
\cite{Dubois2014}. For the sake of the present investigation, we will describe here only the features of interest for the impact of connectivity on the physical properties of the synthetic galaxies.

The \hagn simulation makes use of the adaptive-mesh refinement code \ramses \citep{teyssier02}.
The simulation is run in a box size of $L_{\rm  box}=100 \, h^{-1}\,\mathrm{Mpc}$ with a $\Lambda$CDM cosmology compatible with the 7-years Wilkinson Microwave Anisotropy Probe data \citep{WMAP2011}, with total matter density $\Omega_m = 0.272$, dark energy density $\Omega_\Lambda = 0.728$, baryon density $\Omega_b = 0.045$, Hubble constant $H_0 = 70.4$ km s$^{-1}$ Mpc$^{-1}$, amplitude of the matter power spectrum $\sigma_8 = 0.81$, and $n_s = 0.967$. 
It contains $1024^3$ dark matter particles (i.e. a mass resolution of $M_\mathrm{DM,res}=8 \times 10^7 \, \msun$) and   
the initial grid is refined down to $1$~physical kpc. The refinement is triggered when the number of  particles becomes greater than 8 (or if the total baryonic mass reaches 8 times the initial dark matter mass resolution in a cell).
A uniform UV background is switched on at $z_{\rm  reion} = 10$ \citep{haardt&madau96}. 
Gas cools down to $10^4\, \rm K$ via H, He and metals \citep{sutherland&dopita93}. 
Star particles are created in regions where gas number density reaches $n_0=0.1\, \rm H\, cm^{-3}$,  following a Schmidt relation: $\dot \rho_\star= \epsilon_\star {\rho_{\rm g} / t_{\rm  ff}}$,  with $\dot \rho_\star$  the star formation rate mass density, $\rho_{\rm g}$ the gas mass density, $\epsilon_\star=0.02$ the constant star formation efficiency  and $t_{\rm  ff}$ the gas local free-fall time.
\hagn implements a subgrid feedback from stellar winds and  supernova (both type Ia and  II).
\hagn also follows galactic black hole (BH) formation and active galactic nuclei (AGN) feedback. BHs can grow by gas accretion at a Bondi-Hoyle-Lyttleton rate capped at the Eddington accretion rate when they form a tight enough binary. The AGN feedback is a combination of two different modes, the so-called quasar and radio mode, in which BHs release energy in the form of bipolar jet or heating when the accretion rate is respectively below and above 1\% of Eddington ratio, with efficiencies tuned to match the BH-galaxy scaling relations at $z=0$ \citep[see][for details]{Dubois2012}.

Galaxies are identified from the stellar particles distribution using the \textsc{AdaptaHOP} halo finder \citep{aubert04}. The local stellar particle density is computed from the 20 nearest neighbours, and are kept when they match the density threshold  of 178 times the average matter density (for the  dark matter halos  a density threshold of 80 times the average matter density is applied when they contain  more than 100 particles).  
Each galaxy is then associated with its closest main halo.

In order to assess the impact of AGN feedback on the connectivity of the cosmic web and galaxy properties, the analysis also relies on the \hnoagn simulation. This simulation uses identical initial conditions and sub-grid modelling, but does not model black hole formation, nor AGN feedback \citep{duboisetal16,Peirani17}.

\subsubsection{The SIMBA Meshless cosmological simulation}

The third simulation is the \simba simulation \citep{Dave2019} built on the \mufasa suite~\citep{Dave2016}.
It uses the mass-conserving Meshless Finite Mass version of the \gizmo code \citep{Hopkins2015}. 
\simba follows the evolution of 1024$^3$ dark matter particles and 1024$^3$ gas elements in a volume of 100$^3$ ($h^{-1}$ Mpc$)^3$, with a minimum gravitational softening length of 0.5  $h^{-1}$ comoving kpc, assuming a \cite{Planck2016} $\Lambda$CDM cosmology with  $\Omega_m = 0.3$, $\Omega_\Lambda = 0.7$, $\Omega_b = 0.048$, $H_0 = 68$ km s$^{-1}$ Mpc$^{-1}$, $\sigma_8 = 0.82$, and $n_s = 0.97$. 
The initial gas element mass is 1.82 $ \times 10^{7}$ \msun, and the dark matter particle mass resolution is 9.6 $\times 10^{7}$ \msun.  The volume and resolution are thus quite comparable to \hagn. 
Radiative cooling and photoionisation heating are modeled, including metal cooling and non-equilibrium evolution of primordial elements. A spatially uniform ionising background switched on at $z_{\rm  reion} \sim 10.7$ is also assumed from \citet{haardt&madau2012}, modified to account for self-shielding so that the neutral hydrogen content of gas elements is modeled self-consistently.
Star formation is based on the H$_2$ content of the gas and the H$_2$ fraction computation is based on the metallicity and local column density and follows the sub-grid model of \cite{KrumholzGnedin2011}.
The  star  formation rate is therefore  given  by SFR = $\epsilon_{\star} \rho_\mathrm{H_2}/t_{\rm dyn}$, where $\rho_{\rm{H}_2}$ is the H$_2$ density and $\epsilon_{\star}=0.02$.
Stellar feedback is modeled via two-phase kinetic decoupled galactic winds, in which 30\% of wind particles are ejected "hot".
Black hole growth is accounted for via the torque-limited accretion model \citep{AA2017} from cold gas and Bondi accretion from hot gas. AGN feedback is modelled via kinetic bipolar outflows.  These result in a population of star-forming and quenched galaxies and their black holes that are in good agreement with observations~\citep{Dave2019,Thomas2019}.

Halos are identified on the fly using a 3D Friends-of-Friends (FoF) algorithm within \gizmo, with a linking length that is 0.2 times the mean inter-particle distance. Galaxies are identified using a post-processed 6D FoF galaxy finder. Galaxies and halos are cross-matched and their properties computed using the \yt-based package \caesar\footnote{\href{https://caesar.readthedocs.io/}{caesar.readthedocs.io}}.

\subsubsection{Galaxy kinematics and properties}

As a proxy for morphology, we use for each galaxy the kinematic ratio of their rotation to dispersion velocity, \vsig, computed from the 3D velocity distribution of stars. 
In order to do this, the $z$-axis of the cylindrical spatial coordinates ($r$, $\theta$, $z$) is chosen to be oriented along the total angular momentum (spin) of stellar component of the galaxy.
The velocity of each stellar particle is then decomposed into cylindrical components $v_r$, $v_{\theta}$, $v_z$, and the rotational velocity of a galaxy $v$ is defined as the mean of $v_\theta$ of its individual stars. The average velocity dispersion of the galaxy is computed as $\sigma^2 = (\sigma^2_r + \sigma^2_\theta + \sigma^2_z)/3$ using the velocity dispersion of each velocity component $\sigma_r$, $\sigma_\theta$ and $\sigma_z$. While this is not directly comparable to observed measures of \vsig, it allows us to separate rotation-dominated from dispersion-dominated systems~\citep{duboisetal16,Kraljic2019}.
The SFR of virtual galaxies in \hagn is computed over a time-scale of 50 Myr\footnote{We do not attempt to match exactly this criterion to the SDSS derivation of the SFR where for normal star-forming galaxies this is based on the $H_\alpha$ emission (see Section~\ref{sec:obs-data})
The corresponding time-scale is $\sim$ 10 Myr \citep[e.g.][]{Kennicutt2012}, which is comparable to the one adopted in the simulations. However, we have also tested shorter and longer time-scales, i.e. 10 and 100 Myr in the \hagn simulation and we find that this does not change our conclusions.}, while in \simba, SFR is computed from the gas particles\footnote{We checked that these values are correlated with those obtained from the stellar particles computed over a time-scale betwen 50 to 100 Myr.}. 

We finally stress that our catalogues (observed and virtual) contain both centrals and satellites, and that we expect the physical properties of these sub-populations to reflect their nature \citep[for instance starvation,][will only operate on satellites]{larsonetal1980}.
This effect is investigated in Appendix~\ref{sec:satellite-mass}, in the context of the stellar mass--connectivity relation.

\section{Connectivity and multiplicity in the SDSS}
\label{sec:results}

Let us now characterise the connectivity and multiplicity (or local connectivity, see Section~\ref{sec:cw}) of SDSS galaxies.
In Section~\ref{sec:StellarMass}, we investigate specifically the impact of stellar mass on connectivity while Section~\ref{sec:SSFR} studies the impact of connectivity on the \ssfr at fixed stellar mass.

\begin{figure}
\centering\includegraphics[width=1.\columnwidth]{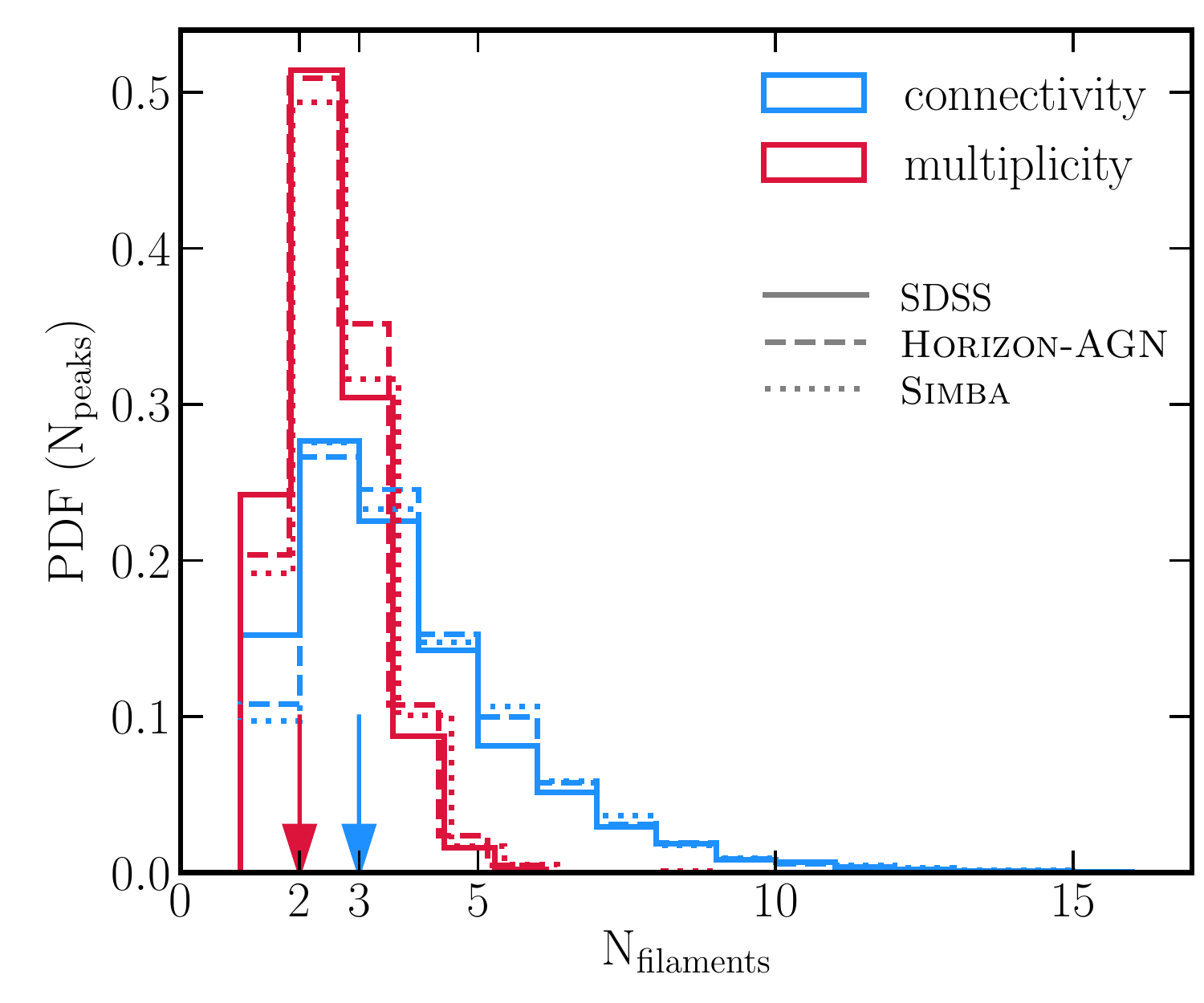}
\caption{PDF of the connectivity (blue) and multiplicity (red) in the SDSS (solid lines), \hagn (dashed lines) and \simba (dotted lines). Arrows show medians of the distributions, 3 for connectivity and 2 for multiplicity, for all three data sets. See Tables~\ref{tab:mean_med_conn} and \ref{tab:mean_med_mult} for mean and median values of connectivity and multiplicity, respectively, as a function of persistence.  The consistency between the SDSS and the simulation is remarkable, illustrating the robustness of such estimators. 
}
\label{fig:connect_multi_pdf}
\end{figure}

Let us start by the distribution of multiplicity and that of connectivity. Figure~\ref{fig:connect_multi_pdf} shows the histograms of connectivity and multiplicity as measured from the full SDSS galaxy distribution in 3D. The medians of these distributions are 3 and 2 for connectivity and multiplicity, respectively. The mean values are fairly similar, $3.26 \pm 0.02 $ and $2.25\pm 0.01$, respectively.  
On this figure, in anticipation of the next section, the measurements from hydrodynamical simulations are also shown as dashed and dotted lines, respectively. 
Two simulations with different numerical schemes and subgrid physics are shown in order to get an idea of the modeling uncertainty. The agreement between real and simulated measurements is quite good.
Note that these mean values are lower than the predictions for Gaussian random fields (resp. 6.1 and 4). This is expected since here, galaxies at all stellar masses are considered and gravitational clustering and dark energy both disconnect the cosmic web, thus decreasing the connectivity \citep{codisetal2018}. With increasing stellar mass (\mstar $\sim 10^{11}-10^{12}$ \msun), connectivity starts to grow towards values for Gaussian random field (see Section~\ref{sec:StellarMass}).
Note also that the low sampling of the field introduces shot noise, which generally tends to disconnect it, as shown on topological estimators like genus by e.g. \cite{Appleby2017}. 
Since we rely on a discrete galaxy distribution from the simulations, this effect {should be}  also accounted for there. 
Given that the PDF is narrower for the multiplicity, 
providing thus less leverage over environment on larger scales,  
we will restrict the rest of the analysis in the main text to the connectivity and refer to Appendix~\ref{sec:multiplicity} for complementary results on multiplicity. 

\subsection{Stellar mass dependence}
\label{sec:StellarMass}

\begin{figure}
\includegraphics[width=1.\columnwidth]{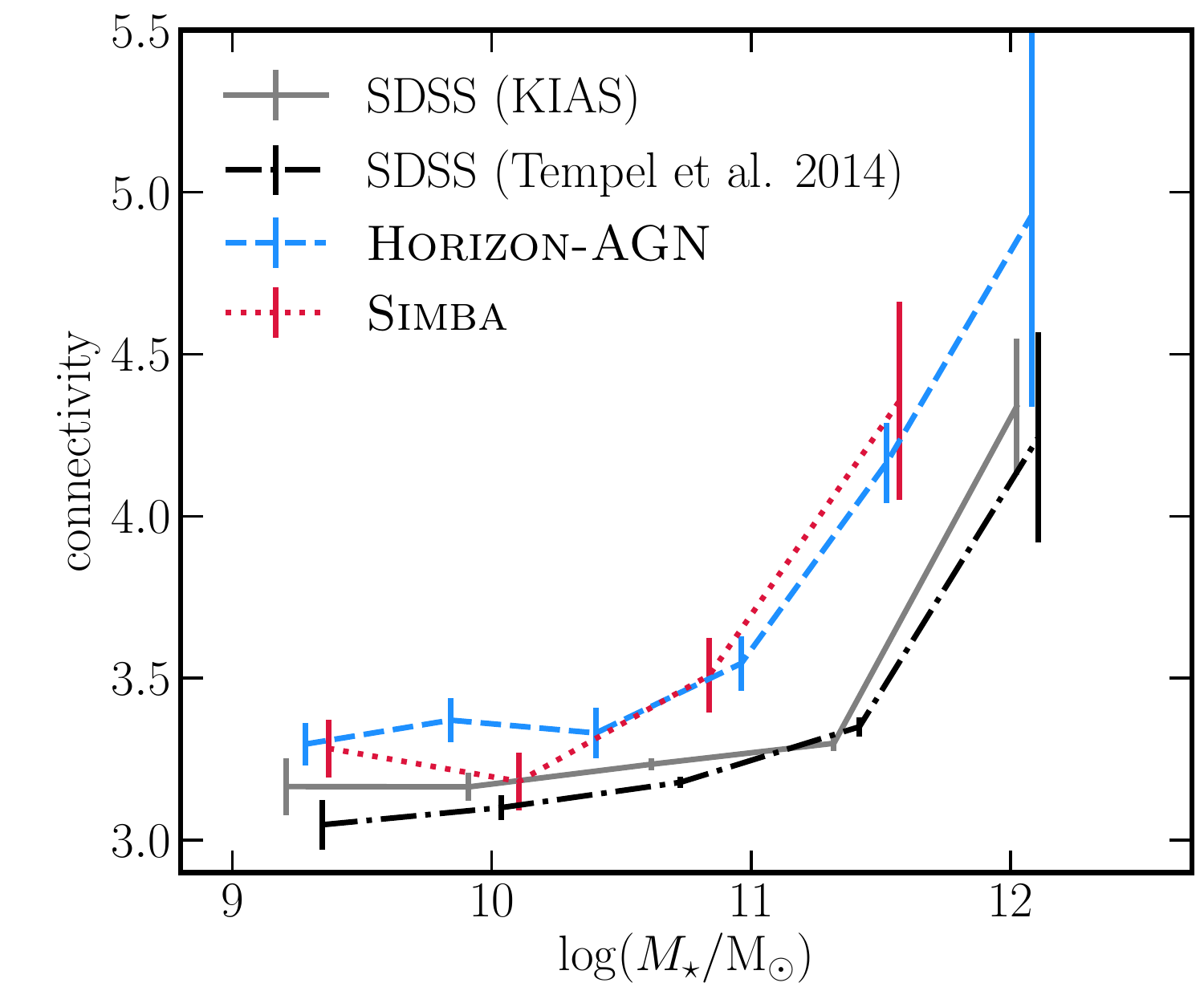}
\caption{Mean connectivity as a function of \mstar in the SDSS (solid grey line), \hagn (dashed blue line) and \simba (dotted red line). In order to quantify our uncertainties, we also use the \protect\cite{Tempel2014} SDSS catalogue (dash-dotted black line). Both sets of simulations and post-processing of the raw SDSS data from both catalogues yield consistent measurements. Connectivity increases with increasing \mstar in qualitative agreement between observations and simulations. The residual difference in amplitude is driven by sampling at fixed persistence, see Appendix~\ref{sec:persistence}.
}
\label{fig:connect_mass_all}
\end{figure}

We now consider the variation of connectivity with stellar mass.  Figure~\ref{fig:connect_mass_all} shows the connectivity 
averaged over galaxies in bins of stellar mass, \mstar, for the entire population (solid grey line). 
More massive galaxies are found to have higher connectivity compared to their lower mass counterparts.
For the sake of quantifying our uncertainties, we also use the \cite{Tempel2014} SDSS catalogue (dash-dotted black line) and find no significant differences.  

The residual difference in the amplitude of the measured connectivity between the simulations and observations is driven by sampling at fixed persistence and the difference in the mean number density in both data sets\footnote{The mean galaxy number density in the SDSS at $z \leq 0.19$ is $4.4 \times 10^{-3}$ Mpc$^{-3}$, while it is $1.3 \times 10^{-1}$ ($h^{-1}$ Mpc)$^{-3}$ and $4.9 \times 10^{-2}$ ($h^{-1}$ Mpc)$^{-3}$ in the \hagn and \simba simulations at $z=0$, respectively.}. Indeed, if the matter density field is better sampled with a more complete catalogue, more filaments will be recovered at a given persistence (see Appendix~\ref{sec:persistence} for more details). 

We also note that the elbow in the connectivity--\mstar relation, below which there is only a weak trend, happening near 5$\times$10$^{11}$~\msun, corresponds to the Press-Schechter mass of non-linearity $M_*(z=0)$ times the baryon fraction, reflecting the transition between fully non-linear satellites and centrals which have collapsed recently (cf Appendix~\ref{sec:satellite-mass}, Figure~\ref{fig:connect_mass_central}).

\begin{figure}
\centering\includegraphics[width=\columnwidth]{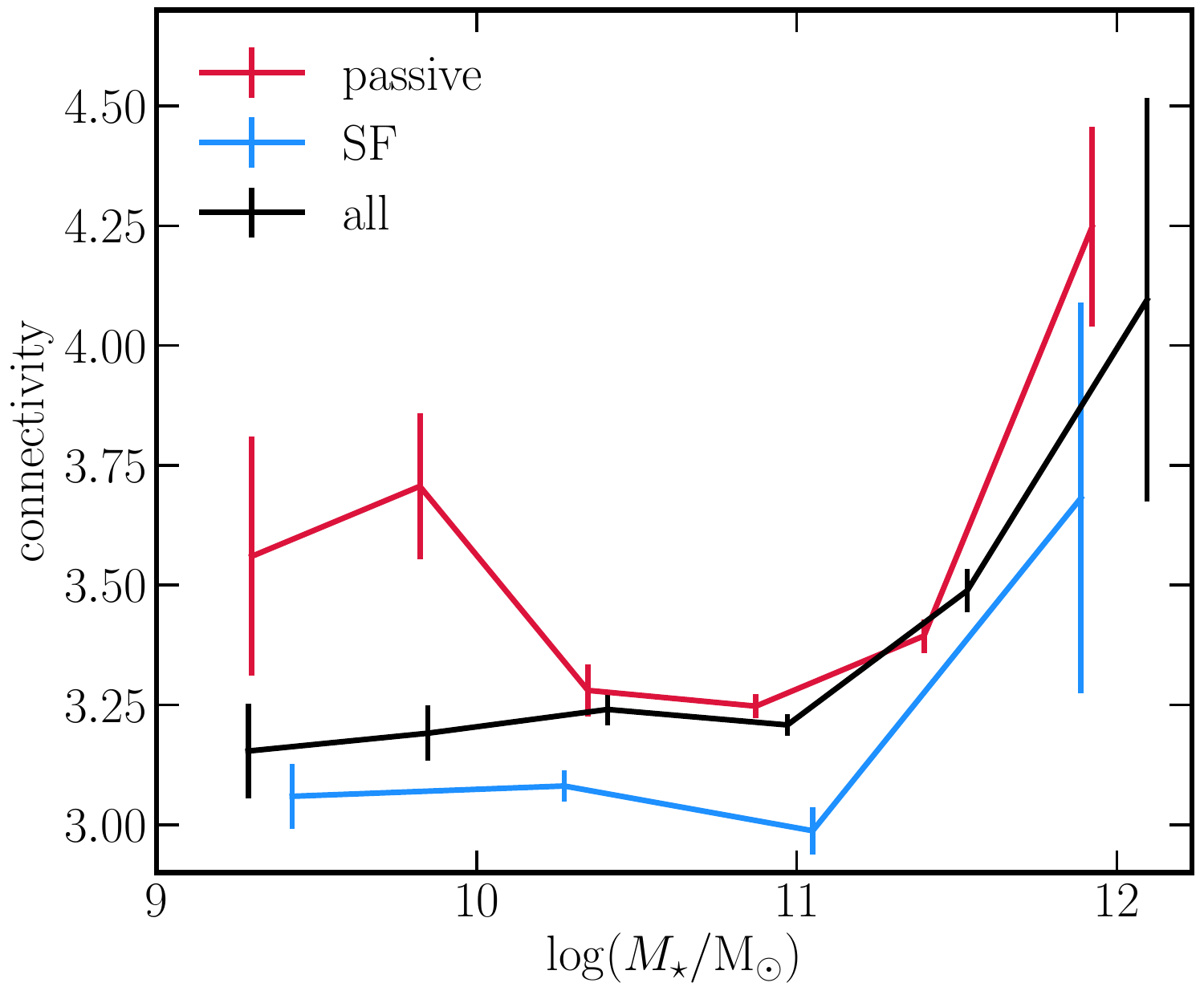}
\centering\includegraphics[width=\columnwidth]{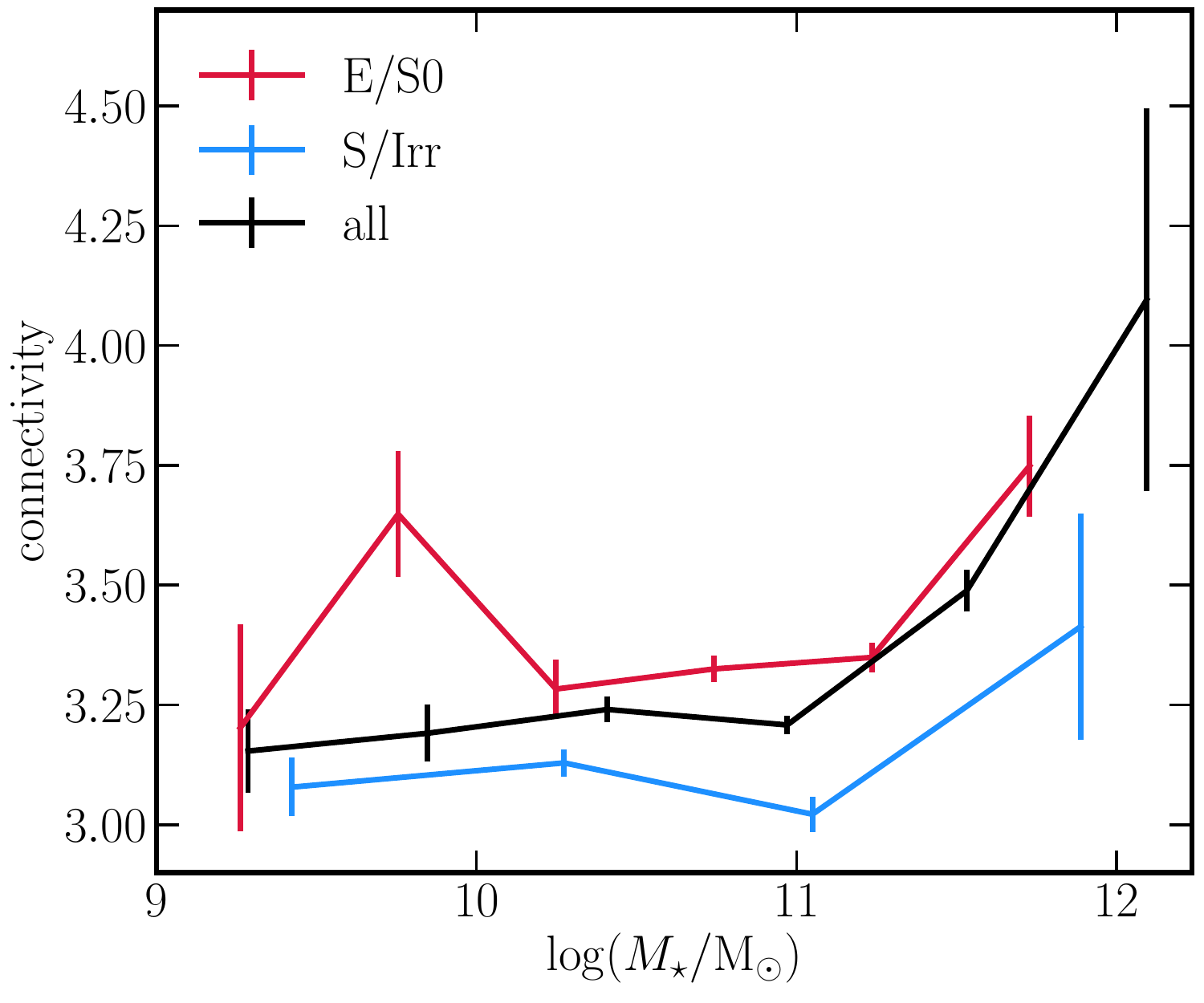}
\caption{Mean connectivity as a function of stellar mass for different galaxy populations in the SDSS. \textsl{Top panel} shows results for the split by star formation activity of galaxies, \textsl{bottom panel} for the split by their morphology. Black solid line in each panel shows the mean connectivity for all galaxies at fixed mass.
More massive galaxies are found to have higher connectivity than their lower stellar mass counterparts. Passive or E/S0 type galaxies (red lines) have higher connectivity than star-forming or S/Irr type galaxies (blue lines) at the same \mstar.
}
\label{fig:connect_mass}
\end{figure}

Figure~\ref{fig:connect_mass} shows the connectivity averaged over galaxies in bins of \mstar for populations with different star formation activity and morphologies. 
In each panel, the black solid line shows the mean connectivity for all galaxies at fixed mass. 
In the top panel of Figure~\ref{fig:connect_mass}, we distinguish between passive and star-forming galaxies (see Section~\ref{sec:obs-data}).
At fixed stellar mass, passive galaxies (red line) tend to have higher connectivity than star forming galaxies (blue line).
In turn, when splitting galaxies by their morphological type (bottom panel), E/S0 galaxies (red line) are found to have higher connectivity than S/Irr galaxies (blue line) of same \mstar.
Although the redshift range is different, this finding is consistent with \cite{DarraghFord2019}, who investigated the link between connectivity, group mass and the properties of the brightest group galaxy at fixed group mass.
Note that for passive/elliptical galaxies, there seems to be a bimodality in the distribution of connectivity and stellar mass as revealed by the non-monotonicity of the red curve on both panels of Figure~\ref{fig:connect_mass}. 
In particular, we tend to see a first bump at low mass and a second rise at larger mass. This is due to the two quite distinct populations of passive ellipticals: low mass satellites on the one hand and massive centrals on the other hand, which may respectively drive the low and high mass behaviours (see Appendix~\ref{sec:satellite-mass}).

\subsection{\ssfr dependence}
\label{sec:SSFR}

It is well-known that the color distribution of galaxies in the local universe is bimodal \citep[e.g.][]{Strateva2001,Baldry2004,Brammer2009}, with red galaxies dominated by massive quiescent spheroids, and blue galaxies being typically lower-mass star-forming disks \citep[e.g.][]{kauffmann2003b,Balogh2004,Moustakas2013}.

\begin{figure}
\centering\includegraphics[width=\columnwidth]{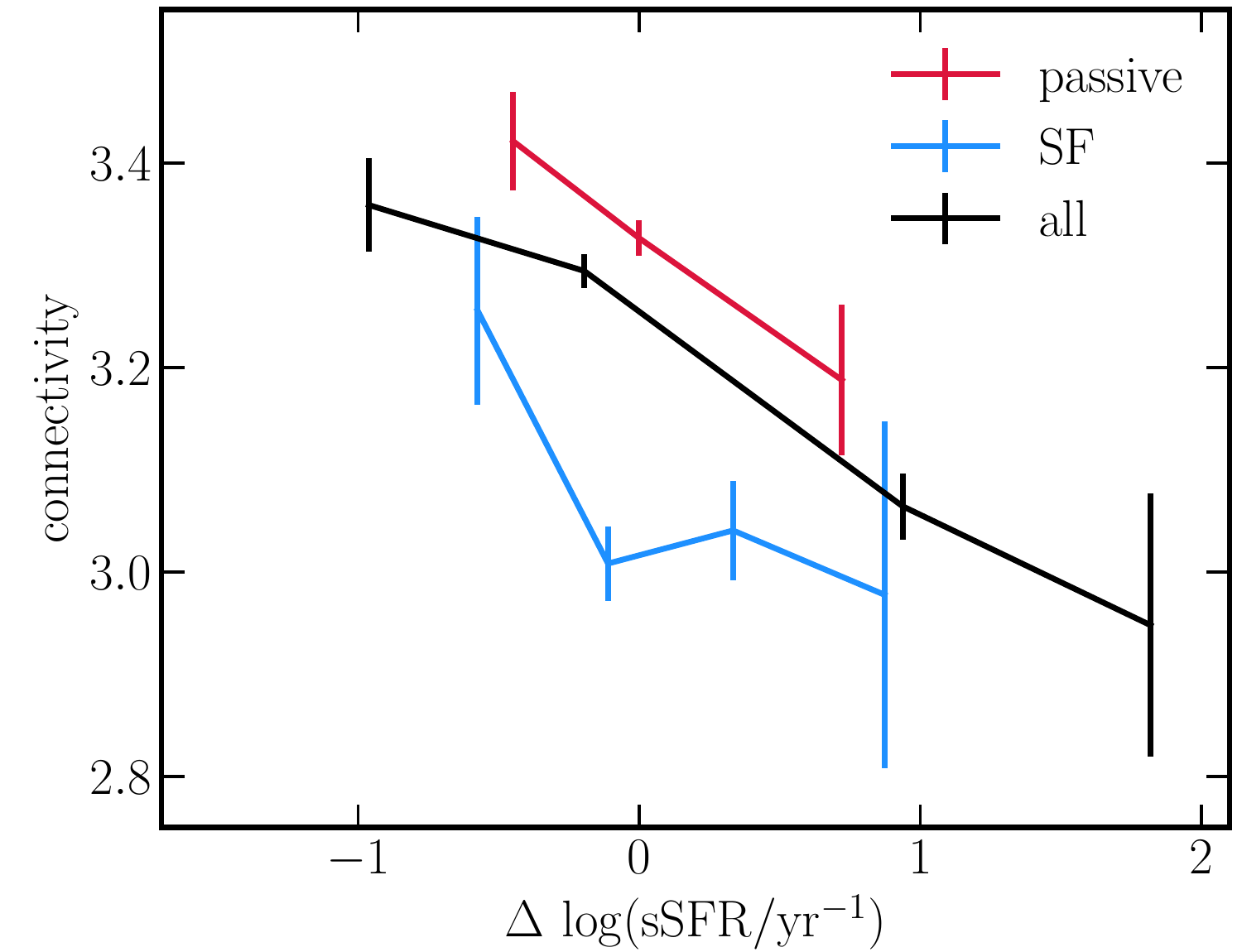}
\centering\includegraphics[width=\columnwidth]{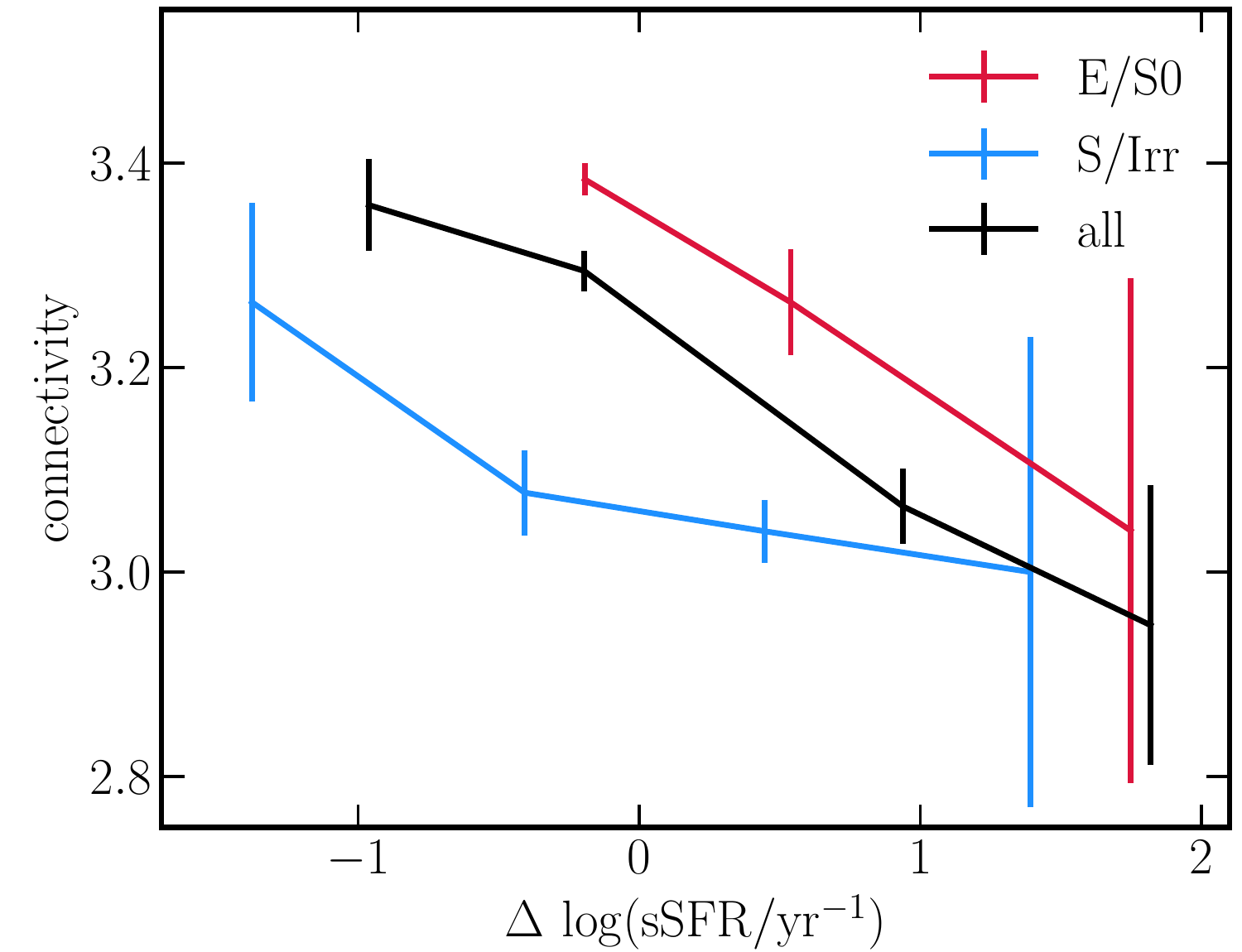}
\caption{Mean connectivity as a function of the excess of \ssfr at given stellar mass for SDSS galaxies. 
\textsl{Top panel} shows results for the split by star formation activity of galaxies, \textsl{bottom panel} for the split by their morphology. Black solid line in each panel shows the mean connectivity for all galaxies. 
Both passive and star-forming galaxies with higher (lower) connectivity tend to have lower (higher) \ssfr than the average at fixed \mstar of a given population. At fixed \ssfr excess, passive galaxies tend to be more connected than star-forming ones.  
Similarly, both elliptical/S0 and spiral/irregular galaxies higher (lower) connectivity tend to have lower (higher) \ssfr than the average at fixed \mstar of a given population. At fixed \ssfr excess, elliptical/S0 galaxies tend to be more connected than galaxies with spiral/irregular morphology.  
}
\label{fig:connect_deltaSSFR}
\end{figure}

In order to account for this color/\ssfr-- or morphology type--\mstar dependence,  Figure~\ref{fig:connect_deltaSSFR} shows the connectivity as a function of excess \ssfr at a given stellar mass, $\Delta \log ({\rm sSFR}/{\rm yr}^{-1})$. The excess is computed at fixed stellar mass with respect to the mean value of a given sub-population of galaxies. 
This allows us to probe the correlation between galaxy \ssfr/type and connectivity beyond what is driven by stellar mass.
We find that galaxies with higher connectivity tend to have lower \ssfr than the average at the same \mstar, while those with lower connectivity have higher \ssfr.

When splitting galaxies by their star formation activity (top panel of Figure~\ref{fig:connect_deltaSSFR}), both passive and star forming galaxies are found to follow qualitatively similar trends for connectivity as a function of the excess \ssfr, compared to what is found for the entire population. In addition, passive galaxies tend to have higher connectivity than star forming galaxies at fixed \ssfr excess.

When splitting galaxies by their morphology type (bottom panel of Figure~\ref{fig:connect_deltaSSFR}), E/S0 galaxies are found to follow a qualitatively similar trends for connectivity as a function of the excess \ssfr to the entire galaxy population, i.e., E/S0 galaxies with reduced \ssfr at fixed \mstar have higher connectivity than E/S0 galaxies showing an excess in \ssfr.
S/Irr galaxies show similar, albeit weaker, dependence in the connectivity -- $\Delta \log ({\rm sSFR}/{\rm yr}^{-1})$ parameter space. 
S/Irr that have lower \ssfr compared to the average at the same \mstar have higher connectivity, compared to those with positive \ssfr excess that are (within the error bars) consistent with no dependence on the number of connected filaments.  
However, S/Irr galaxies tend to have lower connectivity than E/S0 galaxies, whatever their \ssfr excess (apart from the highest $\Delta \ssfr$ bin with low statistics, where the two galaxy populations are within the error bars indistinguishable).
The relative difference between passive and star forming populations could be partly driven by the difference in stellar mass distributions of passive and star forming galaxies, and E/S0 and S/Irr galaxies. Passive galaxies tend to be more massive than star forming galaxies, they are thus expected to have on average higher connectivity compared to their star forming counterparts. Similarly for E/S0 and S/Irr split, if E/S0 are more massive then S/Irr galaxies, the relative difference in connectivity can be driven by underlying stellar mass dependence of connectivity. 

What is more interesting is that once the \ssfr dependence on stellar mass is accounted for, there is still a clear connectivity dependence for all galaxy populations, regardless of their morphology or star formation activity, such that galaxies with a large \ssfr excess tend to have lower connectivity compared to galaxies with a larger \ssfr deficit.

\section{Comparison to simulations}
\label{sec:predictions}

This section presents measurements of connectivity in the \hagn simulation. We note that qualitatively similar results are obtained in the \simba simulation, albeit with larger uncertainties (see Appendix~\ref{sec:simba_results}).
The cosmic web is reconstructed as in the SDSS data, using galaxies as tracers of the matter distribution.
Note that we use the \hagn and \simba simulations as a reference for the measurement of the connectivity in a large-scale "full-physics" experiment and not as a SDSS-like mock catalogue, given the large difference in volume and completeness.
Section~\ref{sec:StellarMass-Simu} recovers the scaling with stellar mass, while Section~\ref{sec:SSFR-Simu}
investigates the effect of \ssfr at fixed mass on connectivity.

As for observational data (Section~\ref{sec:results}), 
we choose 3$\sigma$ as a fiducial value for the persistence threshold used for the cosmic web reconstruction, even though the volume and sampling are quite different in the simulations. 
We have checked that  similar results are obtained when different values are used, though when increasing $N_\sigma$ to 5 and above, the number of galaxies identified as peaks of the cosmic web is dramatically reduced, making a statistical study less reliable. 
The histogram of the connectivity and multiplicity in the \hagn and \simba simulations are shown in Figure~\ref{fig:connect_multi_pdf}, with median values of 3 and 2, respectively, in very good agreement with values measured in observational data.
The mean values are fairly similar, $3.41 \pm 0.03$ and $2.37 \pm 0.02$, for connectivity and multiplicity, respectively in \hagn, and $3.51 \pm 0.06$ and $2.36 \pm 0.03$, for connectivity and multiplicity, respectively in \simba,
again in agreement with measured values in the SDSS. 

The scaling with the threshold value $N_\sigma$ for the cosmic web extraction is qualitatively similar to the one measured in the SDSS in Section~\ref{sec:results} (see Table~\ref{tab:mean_med_conn} in Appendix~\ref{sec:persistence}). 

\subsection{Stellar mass dependence}
\label{sec:StellarMass-Simu}

\begin{figure}
\centering\includegraphics[width=\columnwidth]{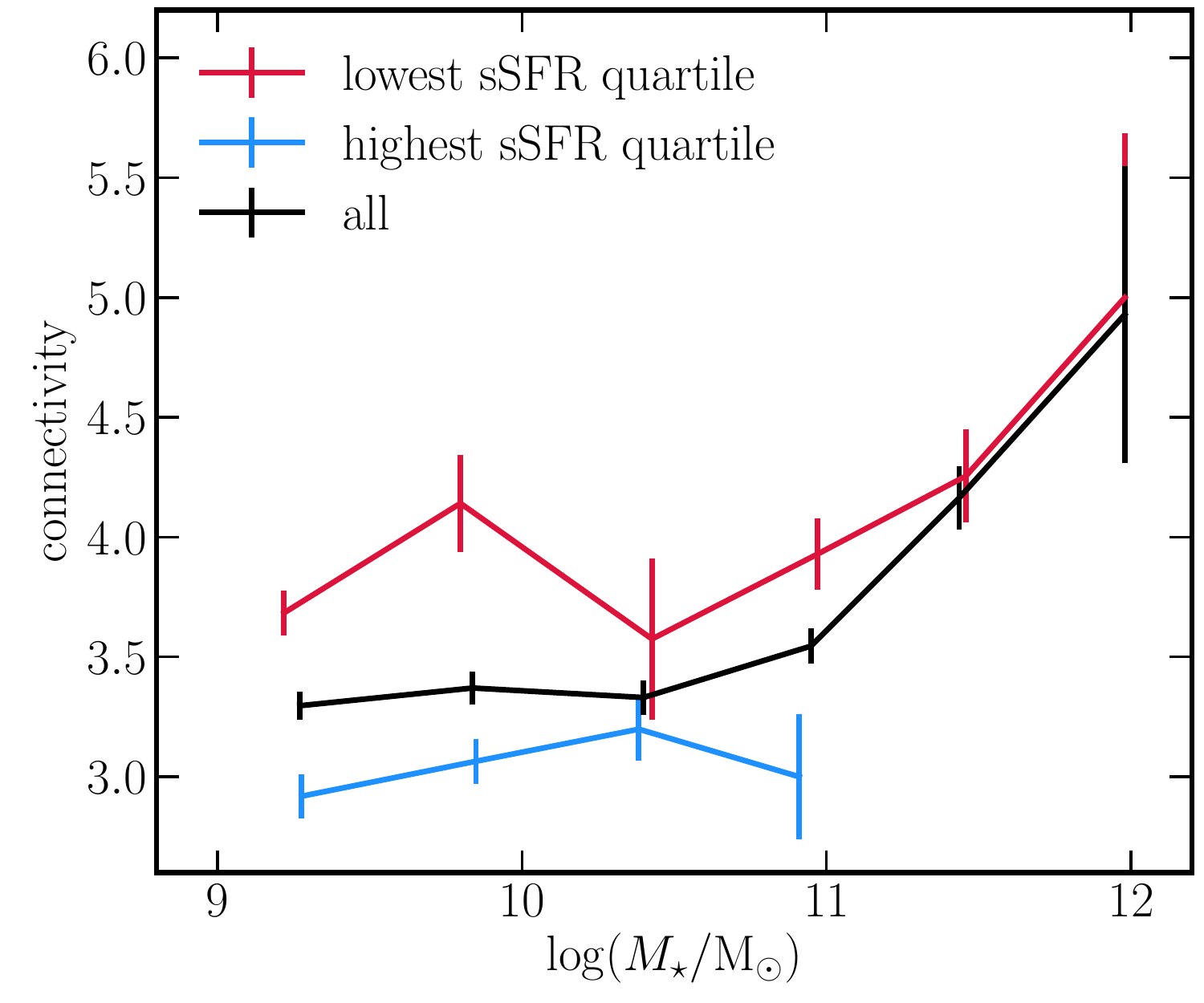}
\centering\includegraphics[width=\columnwidth]{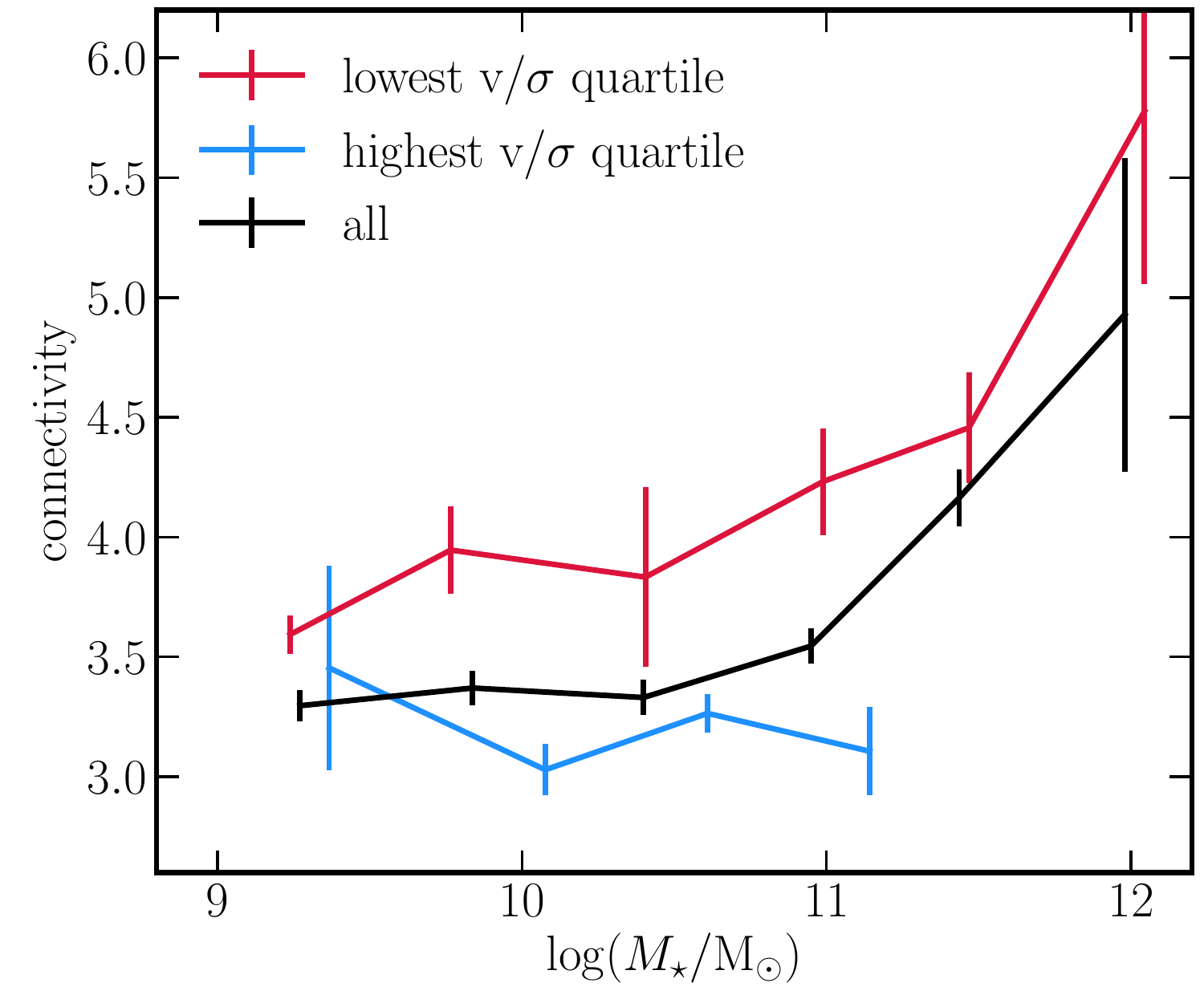}
\caption{Mean connectivity as a function of stellar mass for different galaxy populations in \hagn. \textsl{Top panel} shows results for the split by star formation activity of galaxies, \textsl{bottom panel} for the split by their morphology. Black solid line in each panel shows the mean connectivity for all galaxies at fixed mass.
More massive galaxies are found to have higher connectivity than their lower stellar mass counterparts. 
Galaxies with low \ssfr or ellipticals (red lines) have higher connectivity than star-forming or disk dominated galaxies (blue lines) at the same \mstar, in agreement with trends in the SDSS (see Figure~\ref{fig:connect_mass}). The bump at $\sim 10^{10} \msun$ for galaxies with low \ssfr, also seen in the observations for passive galaxies (see Figure \ref{fig:connect_mass}), corresponds to satellites (see Appendix~\ref{sec:satellite-mass}). 
}
\label{fig:connect_mass_HzAGN}
\end{figure}

Figure~\ref{fig:connect_mass_HzAGN} shows the stellar mass dependence of the connectivity in the \hagn simulation, for populations with different star formation activity and morphologies.
Black lines show connectivity at given stellar mass for the entire galaxy population, reproducing the trends from Figure~\ref{fig:connect_mass_all} and in agreement with observations (see Figure~\ref{fig:connect_mass}).

When splitting by star formation activity (top panel), galaxies with low \ssfr (red line) at fixed stellar mass tend to have higher connectivity than galaxies with high \ssfr (blue line). We note that in the simulation we consider the lowest and highest \ssfr quartile, while in the observational data, we adopted split into passive and star-forming galaxies (see Section~\ref{sec:obs-data}). When we apply the same strategy on the SDSS data, i.e. \ssfr quartiles, we obtain even better agreement with the simulations in particular at high \mstar\footnote{We choose to stick with the passive/star-forming split in the observations for its more common usage in the literature.}.

In turn, when splitting galaxies by their morphological type (bottom panel), parametrised by \vsig, ellipticals (red line) are found to have higher connectivity than disk-dominated galaxies (blue line) of same \mstar. This trend is again in agreement with measurement in observational data (see Figure~\ref{fig:connect_mass}).

\subsection{\ssfr dependence}
\label{sec:SSFR-Simu}

\begin{figure}
\centering\includegraphics[width=\columnwidth]{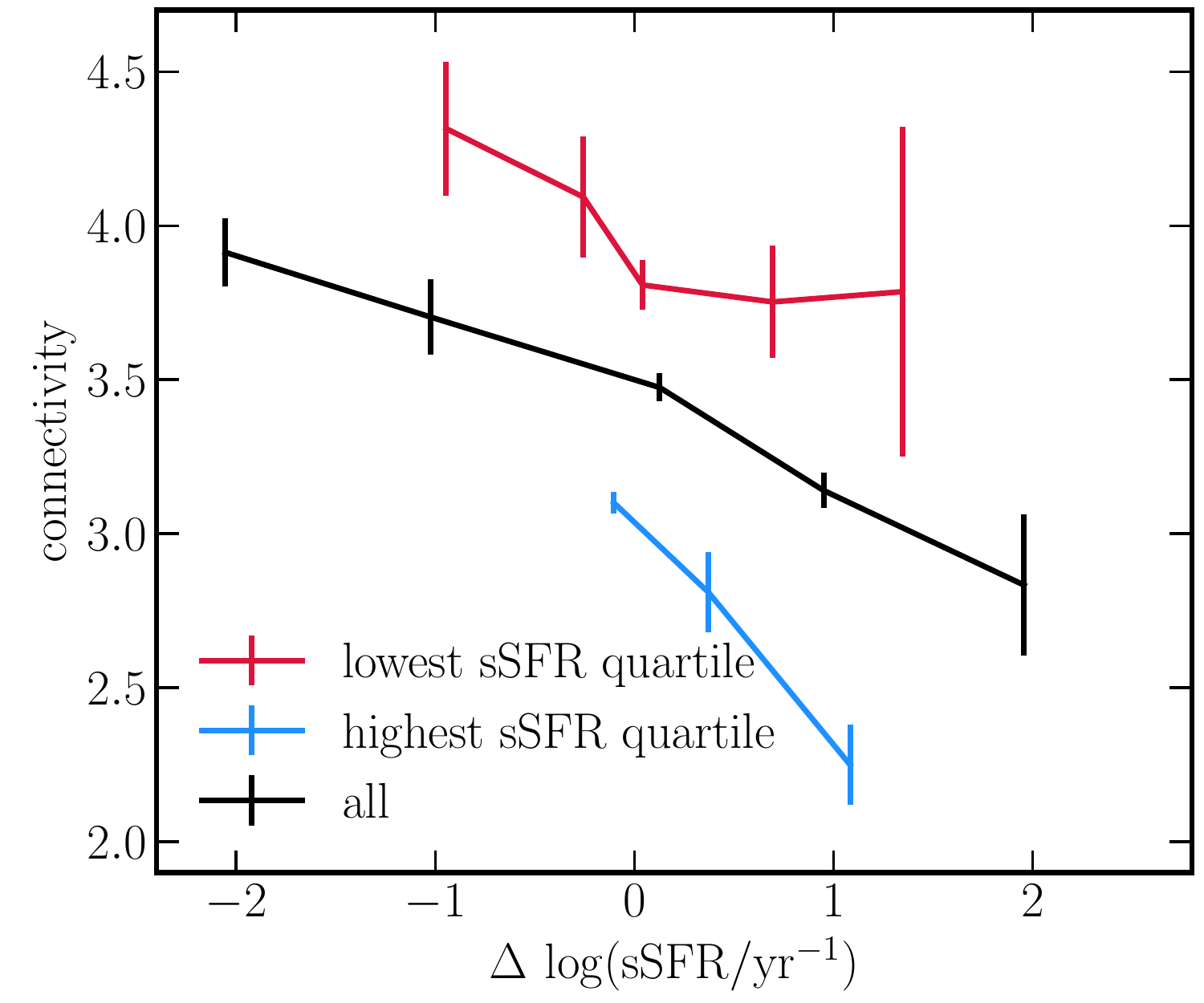}
\centering\includegraphics[width=\columnwidth]{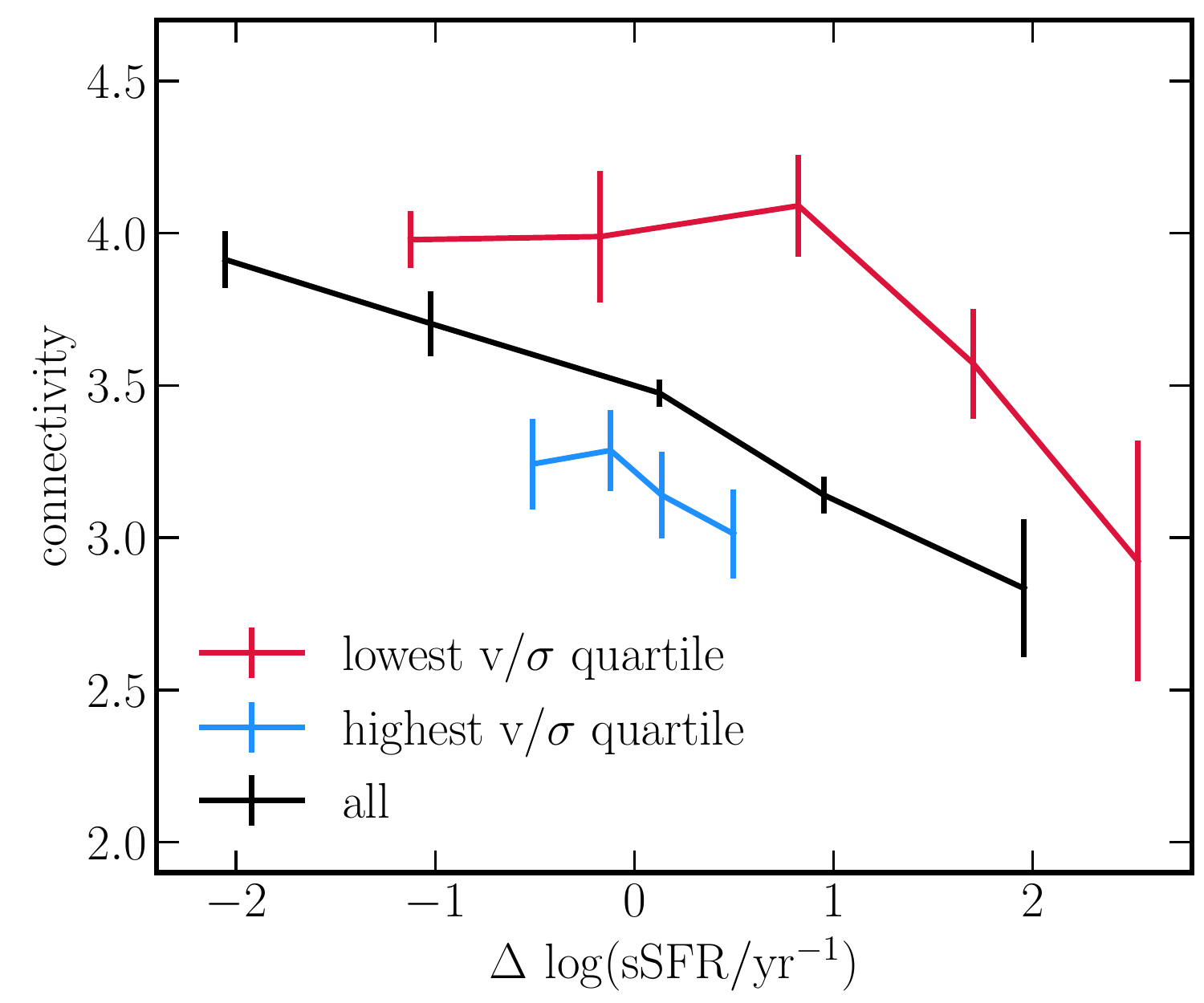}
\caption{Mean connectivity as a function of the excess of \ssfr at given stellar mass in \hagn. 
The \textsl{top panel} is split by star formation activity,  while the\textsl{bottom panel} is split by  morphology. Black solid line in each panel shows the mean connectivity for all galaxies. 
Both passive (red line) and star-forming (blue line) galaxies with higher (lower) connectivity  have lower (higher) \ssfr than the average, at fixed \mstar of a given population. At fixed \ssfr excess, galaxies with low \ssfr are more connected than star-forming ones.  
Similarly, both ellipticals (red line) and disk-dominated (blue line) galaxies with higher (lower) connectivity have lower (higher) \ssfr than the average at fixed \mstar of a given population. At fixed \ssfr excess, ellipticals are more connected than disky galaxies.
Overall, galaxies with higher connectivity have lower \ssfr than the average population at the same \mstar, regardless of their morphology or star formation activity, in agreement with trends seen in observations (see Figure~\ref{fig:connect_deltaSSFR}).
}
\label{fig:connect_deltaSSFR_HzAGN}
\end{figure}

As we did for observations, in order to account for the underlying \ssfr- and galaxy type-\mstar relation, we study connectivity as a function of the excess \ssfr at a given stellar mass.
The result for the entire population of galaxies is shown in 
Figure~\ref{fig:connect_deltaSSFR_HzAGN} (black line).
Galaxies with higher connectivity tend to have lower \ssfr than the average at the same \mstar, while those with lower connectivity have higher \ssfr, once again in agreement with observations (see Figure~\ref{fig:connect_deltaSSFR}).

When splitting galaxies by their star formation activity (top panel of Figure~\ref{fig:connect_deltaSSFR_HzAGN}), both low- (red line) and high-\ssfr (blue line) galaxies are found to follow qualitatively similar trends for connectivity as a function of the excess \ssfr, compared to that is found for the entire population. 
In addition, galaxies with low \ssfr tend to have higher connectivity than galaxies with high \ssfr at fixed \ssfr excess.

When splitting galaxies by morphology (bottom panel of Figure~\ref{fig:connect_deltaSSFR_HzAGN}), ellipticals (red line) are found to follow a qualitatively similar trend in \ssfr vs connectivity  as the entire galaxy population, i.e., elliptical galaxies with reduced \ssfr at fixed \mstar have higher connectivity than ellipticals showing an excess in \ssfr.
Disk-dominated galaxies (blue line) do not show a strong dependence
in the connectivity -- $\Delta \log ({\rm sSFR}/{\rm yr}^{-1})$ parameter space. 
Disky galaxies that have higher \ssfr compared to the average at the same \mstar have lower connectivity, compared to those with negative \ssfr excess that are (within the error bars) consistent with no dependence on the number of connected filaments. Note also that 
contrary to observations, the range of $\Delta \log ({\rm sSFR}/{\rm yr}^{-1})$ values for disk galaxies is tighter. This likely an effect of fairly low resolution that tends to over-smooth the SFR.

However, disk-dominated galaxies tend to have lower connectivity than elliptical galaxies, whatever their \ssfr excess (in the range of values that have in common).

Overall, once the \ssfr dependence on stellar mass is accounted for, there is still a clear connectivity dependence for all galaxy populations, regardless of their morphology or star formation activity, such that galaxies with a large \ssfr excess tend to have lower connectivity compared to galaxies with larger \ssfr deficit, in agreement with the trends found in observations.

\section{Discussion}
\label{sec:discussion}

Let us now re-frame the results of the previous section in terms of the underlying physical parameters, in order to disentangle the known effect of halo mass (effectively the depth of the potential well of the host halo) from the environment (here the connectivity) on galaxy properties (Section~\ref{sec:halo-mass}). We will also quantify the  effect of AGN feedback by comparing the measured connectivity in simulations with and without AGN feedback (Section~\ref{sec:AGNfeedback}). Finally, we will discuss connectivity versus other environmental tracers (Section~\ref{sec:tracer}).

\subsection{Impact of halo mass}
\label{sec:halo-mass}

\begin{figure*}
\centering\includegraphics[width=\textwidth]{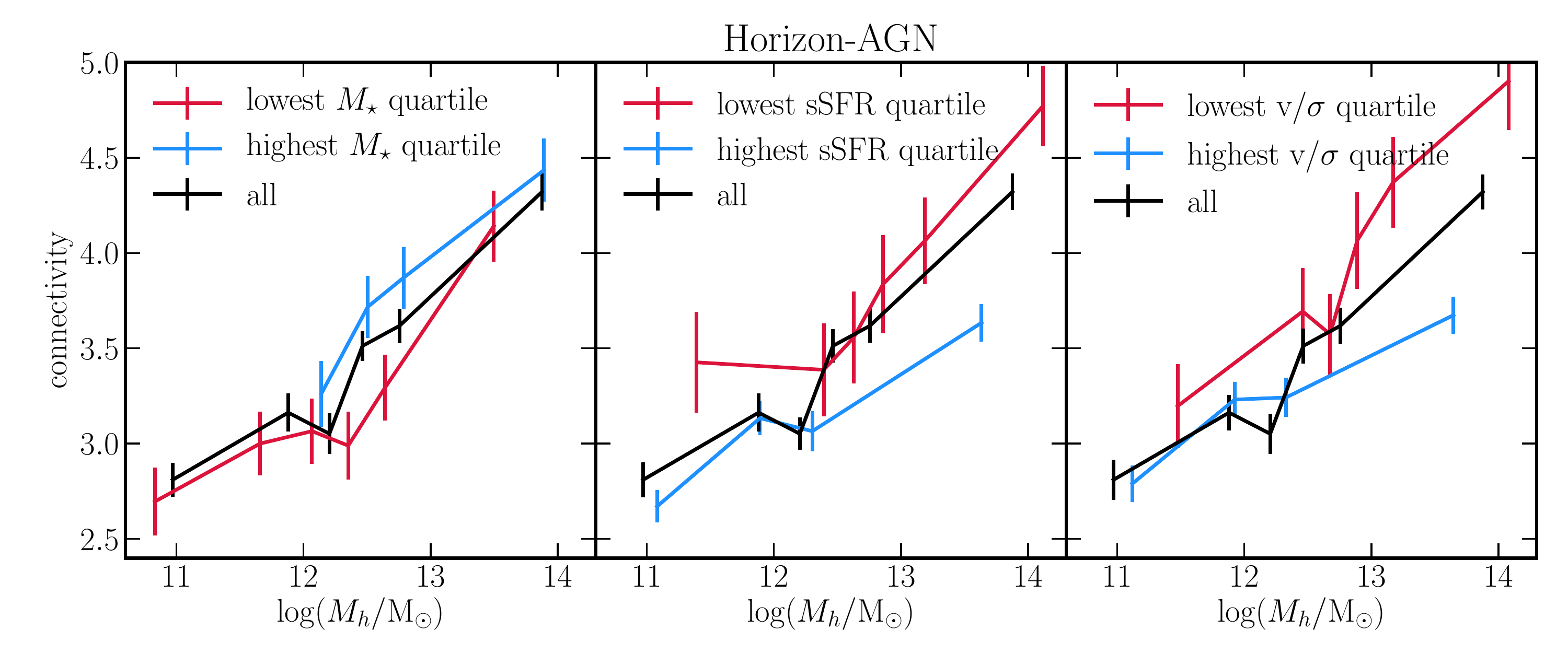}\\
\centering\includegraphics[width=\textwidth]{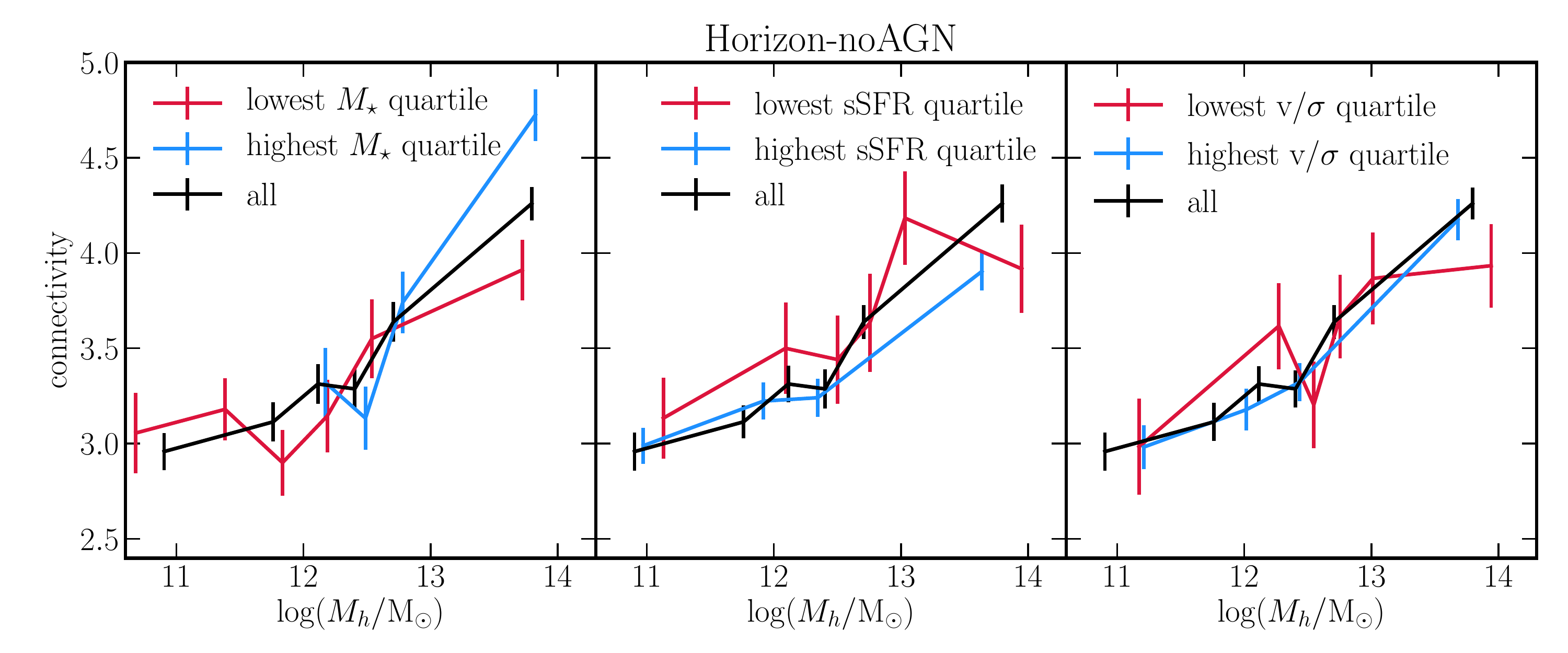}
\caption{Mean connectivity as a function of halo mass in \hagn ({\sl top panels}).  The {\sl left panel} shows results for the split by \mstar, the {\sl middle panel} for the split by star formation activity, and the {\sl right panel} by morphology.
At fixed halo mass, galaxies with higher \mstar, lower \ssfr and lower \vsig tend to have higher connectivity. Beyond halo mass, connectivity directly impacts galactic properties. 
{\sl Bottom panels} show the same quantities for \hnoagn. At all but highest \mhalo, no significant difference in connectivity is seen as a function of explored galaxy properties. At highest halo mass, reversal of the trend compared to \hagn is observed (strongest for \mstar and \vsig).
}
\label{fig:connect_mHalo_3in1}
\end{figure*}

We start by addressing host halo mass, \mhalo, and its impact on the physical properties of galaxies in the framework of connectivity.
Figure~\ref{fig:connect_mHalo_3in1}, top panels, shows the evolution of connectivity as a function of main halo mass
for respectively low and high stellar mass, \ssfr and \vsig in the \hagn simulation. Connectivity increases with increasing halo mass. At fixed \mhalo galaxies with higher \mstar, lower \ssfr and lower \vsig tend to have higher connectivity compared to their lower mass, more star forming and disk dominated counterparts. This clearly shows that properties of galaxies are correlated with connectivity beyond halo mass.
In addition, the trend is stronger for \ssfr and \vsig compared to \mstar suggesting an effect beyond that of halo and stellar mass.

Naively, one could expect, at least at high redshift, that at given halo mass more connected galaxies would be fed by more cold gas, hence be  more massive and star forming. One could also argue that a galaxy embedded in a single filament would be fed more coherent angular momentum hence be more disc dominated. At lower redshift, the net effect of higher connectivity is less obvious since cosmic filaments may not reach down to galaxies, while less frequent minor mergers {along the connected filaments} may be dryer. 
In practice, at high mass the opposite trend is observed at $z=0$ (top panels of Figure~\ref{fig:connect_mHalo_3in1}), i.e. more connected galaxies tend to have lower \ssfr and \vsig.\footnote{We note that we do not separate here between the centrals and satellites, which can be seen on the top-left panel where the  difference at fixed high \mhalo is driven by the population of satellites. For the two remaining properties (\ssfr and \vsig), splitting between centrals and satellites yields qualitatively similar results as for the entire population of galaxies.} 

\subsection{Impact of AGN feedback}
\label{sec:AGNfeedback}

Relying on the companion simulation \hnoagn, the origin of this reversal can be pinned down in part on the impact of AGN feedback. 
As can be seen on bottom panels of Figure~\ref{fig:connect_mHalo_3in1}, in the absence of AGN feedback, at low halo mass ($\mhalo \lesssim 10^{13} \msun$), the connectivity of lowest and highest \ssfr quartiles on the one hand, and (to a lesser extent) lowest and highest \vsig quartiles on the other hand are consistent within the error bars. However, at high halo mass, the highest \mstar quartile displays significantly higher connectivity than the lowest \mstar quartile. We suggest that this is a consequence of the past effect of connectivity on building-up the stellar mass through gas accretion. Such trend is not seen on the \ssfr, which is a more instantaneous quantity, because at low redshift additional quenching processes (e.g. gas shock heating) are likely to overcome the positive impact of connectivity. 
When AGN feedback is included however, there is a clear correlation between high connectivity and passive/elliptical type. At high redshift, one could indeed expect that the higher gas mass load in more connected galaxies might boost the AGN activity, which in turn significantly contributes to quenching and morphological change of galaxies \citep[e.g.][]{Dubois2013,duboisetal16,Pontzen2017}. Indeed, for centrals, AGN feedback quenches star formation by preventing disc to reform. For satellites, it heats their host halo and contributes to starvation of their cold gas, hence quenching star formation \citep[][]{Dashyan2019}. 
In this scenario, the residual dispersion in physical properties at fixed halo mass is driven by connectivity while the ordering (higher connectivity yields higher stellar mass, lower \ssfr and lower \vsig) is driven by AGN feedback at the high mass end. 

Although this interpretation is appealing, one could also speculate that there is no causation between connectivity and important quenching/morphological changes, but that both are driven by the same external cause. 
Interestingly, our results are qualitatively consistent with the higher redshift findings of \cite{DarraghFord2019}, whereby more massive groups have higher {\sl multiplicity}. From the analysis of the \hagn lightcone, they found that AGN feedback quenching efficiency is higher in more connected groups and that the highest-multiplicity groups at given group mass are very likely to be the result of a recent major merger. They suggested that major merger, while increasing the net connectivity of the group (as long as each progenitor has a connectivity higher than 2), might also enhance black hole growth, boosting the AGN feedback strength and therefore quenching the galaxy as a whole \citep[e.g.][]{dimatteoetal05,Springel2005,duboisetal16}. 

In fact it is likely that both  above described effects are at play. They could possibly be disentangled by looking at the redshift evolution of the trend, since the respective importance of mergers and smooth accretion in building-up massive halo mass will be a function of redshift. In order to verify that higher connectivity brings more cold gas to the galaxy, one could in particular reproduce the same measurement in \hnoagn at $z>2$, where cold accretion is supposed to be very efficient. We will address this in future work.

As an final note, the qualitative agreement between our work and \cite{DarraghFord2019} implies that in this regime, the global connectivity captures well enough local multiplicity around groups. It also suggests that there is no significant change in connectivity between redshift zero and one.

\subsection{Connectivity  versus other environment tracers}
\label{sec:tracer}

Let us briefly discuss the pros and cons of connectivity compared to traditional environment tracers.
Density corresponds to number counts  divided by the cube of some given scale, $L$.  Connectivity naturally defines $L$ as the typical distance from nodes  to saddle points along the cosmic web. As such, connectivity is parameter free for a given skeleton\footnote{Connectivity does depend on the persistence level chosen and on the sparsity of the catalogue  -- see Appendix~\ref{sec:persistence}, but the level of persistence should  in practice be  effectively set near $N_\sigma = 3$ -- unless one is attempting to match catalogues of distinct completeness.}.  Connectivity traces specifically matter along filaments, where both the gas and satellite galaxies flow, whereas classical estimators, such as the 5$^{th}$ nearest neighbour, lose track of loci in the large scale  structures ({the density ranges covered by walls and filaments, or filaments and nodes, overlap to some extent}) and are only concerned with the isotropic effect of the environment. Conversely, the number of saddle points paired to a given maximum is determined by the topology of the local cosmic web. Connectivity is therefore a topologically robust\footnote{It is strictly invariant  for any monotonic local transformation of the field. It is also invariant w.r.t. continuous deformations, such as stretching, twisting, crumpling and bending. It is also statistically robust, because it is fairly insensitive to shot noise, which impacts more strongly e.g. nearest neighbours than the number of saddle points \citep{sousbie112}.} estimate which captures the coherence of inflow along preferred directions. 
The robustness of connectivity is to be contrasted to the many existing  definitions of the (isotropically averaged) local density. 
Formally, the dynamical state of a galaxy, which impacts directly its kinematics and morphology, reflects the past and present tides it was subjected to (in its past lightcone), together with the baryonic processes operating within. Since gas shocks iso-thermally in many galactic circumstances,  density ridges are paramount in defining both these tides and the loci of cold flow accretion. Connectivity is  probably the simplest and most straightforward topological quantity capturing the effect of such ridges.

Note finally that this paper focused on the environment of galaxies at the \textit{nodes} of the cosmic web, from which several filaments are branching out. Alternative metrics one could rely on in studying galaxy properties could be the distance to the closest filament \cite[e.g.][]{alpaslanetal2016,chenetal2017,Kleiner2017,Malavasi2017,Poudel2017,Kraljic2018,Laigle2018,CroneOdekon2018}, distance to the closest wall \citep{Kraljic2018} or 3D neighborhood of filaments in the frame of the saddle points \citep{Kraljic2019}. 
These studies typically also point towards an efficient mechanism which quenches star formation and transforms central galaxy morphology from late to early types.

\section{Conclusions}
\label{sec:conclusions}

We investigated the impact of the connectivity of the cosmic web  on the properties of galaxies using the SDSS (via the MPA and KIAS value added catalogues) and confronted those measurements  to predictions from cosmological hydrodynamical simulations \hagn, \hnoagn and \simba, while disentangling the effect of environment, AGN feedback  and mass. Our main results are:
\begin{itemize}
\item The stellar mass dependence of connectivity and multiplicity is in qualitative agreement with theoretical predictions and measurements in dark matter simulations, once we account for the population of satellites. 
\item The stellar mass, star formation activity, and morphology of galaxies all show some dependence on the connectivity (and multiplicity) of the cosmic web. More massive, less star forming, and less rotation supported galaxies tend to have higher connectivity. 
\item These results qualitatively hold both for observed SDSS galaxies and \hagn or \simba virtual galaxies. In the simulations, the connectivity at fixed  halo mass is higher for quenched, low \vsig, more massive galaxies, which suggests that increasing the number of connected filaments reduces star formation and coherent angular momentum acquisition. This likely originates from  filamentary-driven AGN activity, which quenches star formation and prevents disc reformation.
\item The publicly available code {\sc Disperse} provides a flexible, robust and physically motivated tool (via the connectivity) to quantify  the observed geometry of galactic (anisotropic) environment.
\end{itemize}

This work underlines the importance of the anisotropic large-scale environment -- traced by connectivity -- in modulating galaxy properties beyond halo mass.

Future large-field photometric surveys, such as Euclid \citep[][]{Euclid}, HSC \citep{HSC2018...70S...4A} and LSST \citep[][]{LSST}  will be able to confirm and extend these results by probing a wider group mass range and a larger variety of environment (though in 2D) while relying on  state-of-the art photometric redshift extraction techniques \citep[e.g.][]{davidzonetal19,pasquet19}.

Beyond the scope of this paper, it would be of interest to investigate the mass load  and the wetness of the accretion {\sl per filament} (as dry minor mergers  would not contribute to the \ssfr). One should also study how angular momentum is advected as a function of filament strength \citep{pichonetal11}, or consider the importance of the percolated  geometry of the  hot bubbles, as it will likely impact quenching of satellites in the vicinity of nodes. One should also study the cosmic evolution of connectivity with redshift\footnote{Change in connectivity is a measure of a significant transformation of the field, which, as expected, also impacts properties of galaxies which integrate past and present accretion.}. The redshift range of study can be broaden  through Lyman-$\alpha$ tomography \citep[e.g.][ to probe $2<z<3.5$]{Lee2014} or intensity mapping at higher redshifts \citep[e.g.][]{Chang2010}, 
following individual halos, or for the whole population at a given redshift, as the level of accretion  required to trigger star formation activity is very redshift dependent. This will be the topic of future work.

\section*{Acknowledgements}
SC is partially supported by a research grant from Fondation MERAC. CL and JD acknowledge funding support from Adrian Beecroft and the STFC. We thank S. Rouberol for smoothly running the HORIZON cluster for us and T. Sousbie for his help with {\sc Disperse} which is available at the following \href{http://www.iap.fr/users/sousbie/disperse/}{url}. We thank Stephen Appleby for fruitful comments and discussions while this work was carried out. We also thank Daniel Angl\'es-Alc\'azar and Desika Narayanan for helpful discussions.
\simba was run on the DiRAC@Durham facility managed by the Institute for Computational Cosmology on behalf of the STFC DiRAC HPC Facility. The equipment was funded by BEIS capital funding via STFC capital grants ST/P002293/1, ST/R002371/1 and ST/S002502/1, Durham University and STFC operations grant ST/R000832/1. DiRAC is part of the National e-Infrastructure. 
Funding for the SDSS and SDSS-II has been provided by the Alfred P. Sloan  Foundation, the Participating Institutions, the National Science  Foundation, the U.S. Department of Energy, the National Aeronautics and  Space Administration, the Japanese Monbukagakusho, the Max Planck  Society, and the Higher Education Funding Council for England. The SDSS Web Site is http://www.sdss.org/. The SDSS is managed by the Astrophysical Research Consortium for the Participating Institutions. The Participating Institutions are the  American Museum of Natural History, Astrophysical Institute Potsdam,  University of Basel, Cambridge University, Case Western Reserve University,  University of Chicago, Drexel University, Fermilab, the Institute for  Advanced Study, the Japan Participation Group, Johns Hopkins University,  the Joint Institute for Nuclear Astrophysics, the Kavli Institute for Particle Astrophysics and Cosmology, the Korean Scientist Group, the  Chinese Academy of Sciences (LAMOST), Los Alamos National Laboratory, the Max-Planck-Institute for Astronomy (MPIA), the Max-Planck-Institute  for Astrophysics (MPA), New Mexico State University, Ohio State University,  University of Pittsburgh, University of Portsmouth, Princeton University, the United States Naval Observatory, and the University of Washington.


\bibliographystyle{mnras}
\bibliography{author} 



\appendix
\section{Impact of persistence}
\label{sec:persistence}

Let us investigate briefly the impact of varying the persistence on connectivity. Recall that persistence controls the relative density of critical topologically-paired points connected by the skeleton. As such, dropping lower persistence pairs insures that only the most robust filaments are retained, should one want to focus on those. Since \disperse operates directly on the galaxy catalogues, their sparsity also impacts the scale at which filaments can be robustly extracted. Varying the level of persistence of the more complete catalogue provides means to match the less complete one.   

\begin{figure}
\centering\includegraphics[width=1.\columnwidth]{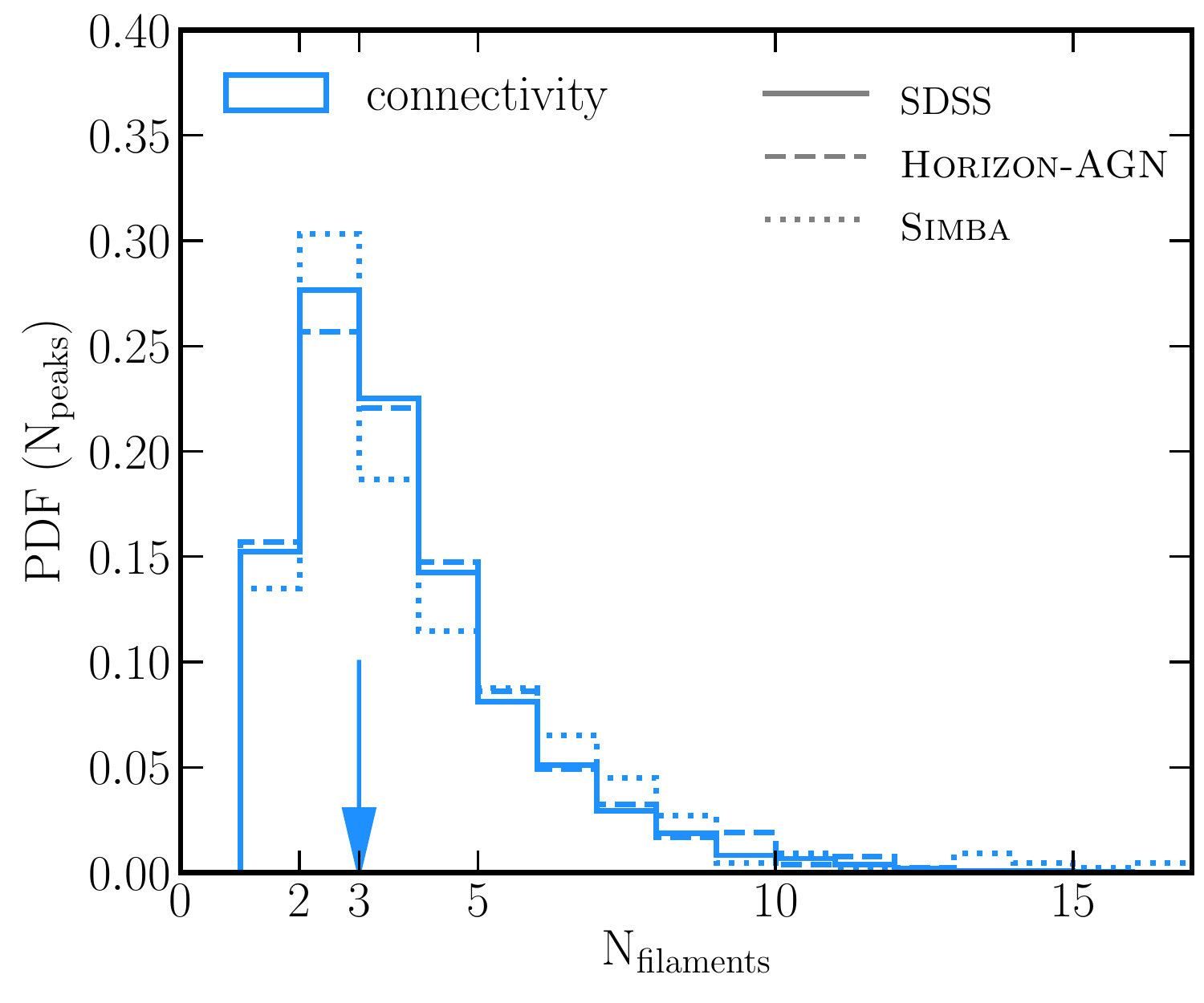}
\caption{PDF of the connectivity in the SDSS (solid) at the persistence level $N_\sigma$ = 3, \hagn (dashed) and \simba (dotted) at the persistence level $N_\sigma$ = 5. The arrow shows median of the distributions for all three data sets. Increased level of persistence in the simulations yields a better agreement with the observations, as it matches better the sampling of the observational data set.  
}
\label{fig:PDF_connect_all_S3_S5}
\end{figure}

\begin{figure}
\centering\includegraphics[width=1.\columnwidth]{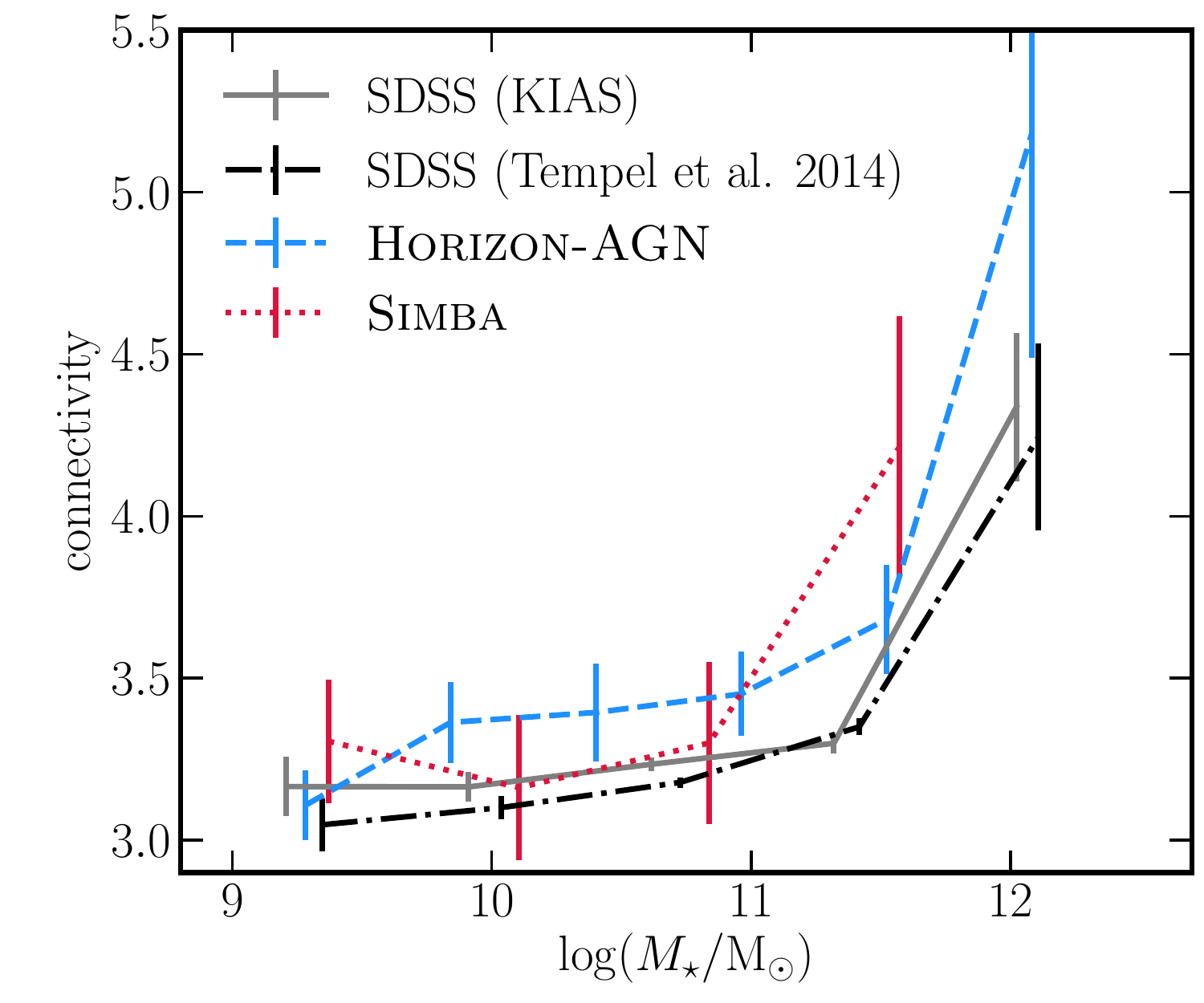}
\caption{Mean connectivity as a function of \mstar in the SDSS (solid grey) at the persistence level $N_\sigma$ = 3, and in \hagn (dashed blue) and \simba (dotted red) at the persistence level $N_\sigma$ = 5. 
Increased persistence in the simulations yields a better agreement with the observations at the expense of increased error bars.  
}
\label{fig:connect_mass_all_S3_S5}
\end{figure}

Figure~\ref{fig:PDF_connect_all_S3_S5} reproduces the PDF  of connectivity  shown in Figure~\ref{fig:connect_multi_pdf}  for a  level of persistence of $5\, N_\sigma$, in order to reflect the fact that the  galactic sampling in the SDSS is much sparser than in the simulations. As expected the agreement between the simulations and the observations is improved at that level of persistence, in particular for the low connectivity bin. 
Similarly Figure~\ref{fig:connect_mass_all_S3_S5} reproduces Figure~\ref{fig:connect_mass_all} for both simulations and the SDSS catalogue.  
One could in principle calibrate more precisely the persistence level to reflect  the difference in  density in both data sets.   
However, in the main text we chose a lower persistence level of $3\, N_\sigma$ in order to avoid having too large error bars, given the much smaller volume of the simulation. We refer to \cite{codisetal2018} for a more detailed investigation of the impact of persistence on connectivity and multiplicity.

Alternatively, one could create a subsample of galaxies from the simulations by matching the mean density in the SDSS. However, such a strategy is not feasible because of the small volume of the simulation. It would lead to a reduction of the number of galaxies by a factor of $\sim$ 10, which would severely impact the cosmic web reconstruction. Additionally, the overall statistics would become insufficient for the current analysis.

\begin{table}
\centering
\begin{threeparttable}
\caption{Mean, standard deviation, 20$^{\rm th}$, 50$^{\rm th}$ and 80$^{\rm th}$ percentiles for connectivity at different persistence levels $N_\sigma$ in the SDSS, and simulations \hagn, \hnoagn and \simba.}
\label{tab:mean_med_conn}
\begin{tabular*}{0.9\columnwidth}{@{\extracolsep{\fill}}lcccccc}
\hline
\hline
\multirow{2}{*}{} & \multirow{2}{*}{$N_\sigma$} & \multirow{2}{*}{mean} & \multirow{2}{*}{std} 
& \multicolumn{3}{c}{percentile}\\
&  &  &  & median &20 & 80\\
\hline
\hline
\multirow{3}{*}{SDSS} & 1 & 4 & 2.2 & 4 & 2 & 6\\
& 3 & 3.3 & 2.0 & 3 & 2 & 5\\
& 5 & 3.3 & 2.0 & 3 & 2 & 5\\
\hline
\multirow{3}{*}{\hagn} & 1 & 4 & 2.2 & 4.1 & 2 & 6\\
& 3 & 3.4 & 1.9 & 3 & 2 & 5\\
& 5 & 3.4 & 2.1 & 3 & 2 & 5\\
\hline
\multirow{3}{*}{\hnoagn} & 1 & 4 & 2.2 & 4 & 2 & 5\\
& 3 & 3.4 & 1.9 & 3 & 2 & 5\\
& 5 & 3.4 & 2.2 & 3 & 2 & 5\\
\hline
\multirow{3}{*}{\simba} & 1 & 4 & 2.2 & 4.1 & 2 & 6\\
& 3 & 3.5 & 2.0 & 3 & 2 & 5 \\
& 5 & 3.6 & 2.5 & 3 & 2 & 5 \\
\hline
\hline
\end{tabular*}
\end{threeparttable}
\end{table}


\begin{table}
\centering
\begin{threeparttable}
\caption{Mean, standard deviation, 20$^{\rm th}$, 50$^{\rm th}$ and 80$^{\rm th}$ percentiles for multiplicity at different persistence levels $N_\sigma$ in the SDSS, and simulations \hagn, \hnoagn and \simba.}
\label{tab:mean_med_mult}
\begin{tabular*}{0.9\columnwidth}{@{\extracolsep{\fill}}lcccccc}
\hline
\hline
\multirow{2}{*}{} & \multirow{2}{*}{$N_\sigma$} & \multirow{2}{*}{mean} & \multirow{2}{*}{std} 
& \multicolumn{3}{c}{percentile}\\
&  &  &  & median &20 & 80\\
\hline
\hline
\multirow{3}{*}{SDSS} & 1 & 3.3 & 1.3 & 3 & 2 & 4 \\
& 3 & 2.2 & 0.9 & 2 & 1 & 3 \\
& 5 & 1.7 & 0.7 & 2 & 1 & 2 \\
\hline
\multirow{3}{*}{\hagn} & 1 & 3.1 & 1.3 & 3 & 2 & 4\\
& 3 & 2.4 & 0.96 & 2 & 2 & 3\\
& 5 & 2.1 & 0.8 & 2 & 1 & 3\\
\hline
\multirow{3}{*}{\hnoagn} & 1 & 3.1 & 1.3 & 3 & 2 & 4\\
& 3 & 2.4 & 0.95 & 2 & 2 & 3\\
& 5 & 2 & 0.81 & 2 & 1 & 3\\
\hline
\multirow{3}{*}{\simba} & 1 & 3.1 & 1.3 & 3 & 2 & 4\\
& 3 & 2.4 & 0.96 & 2 & 2 & 3 \\
& 5 & 1.96 & 0.8 & 2 & 1 & 3 \\
\hline
\hline
\end{tabular*}
\end{threeparttable}
\end{table}

\section{Impact of satellites}
\label{sec:satellite-mass}

\begin{figure}
\centering\includegraphics[width=1.\columnwidth]{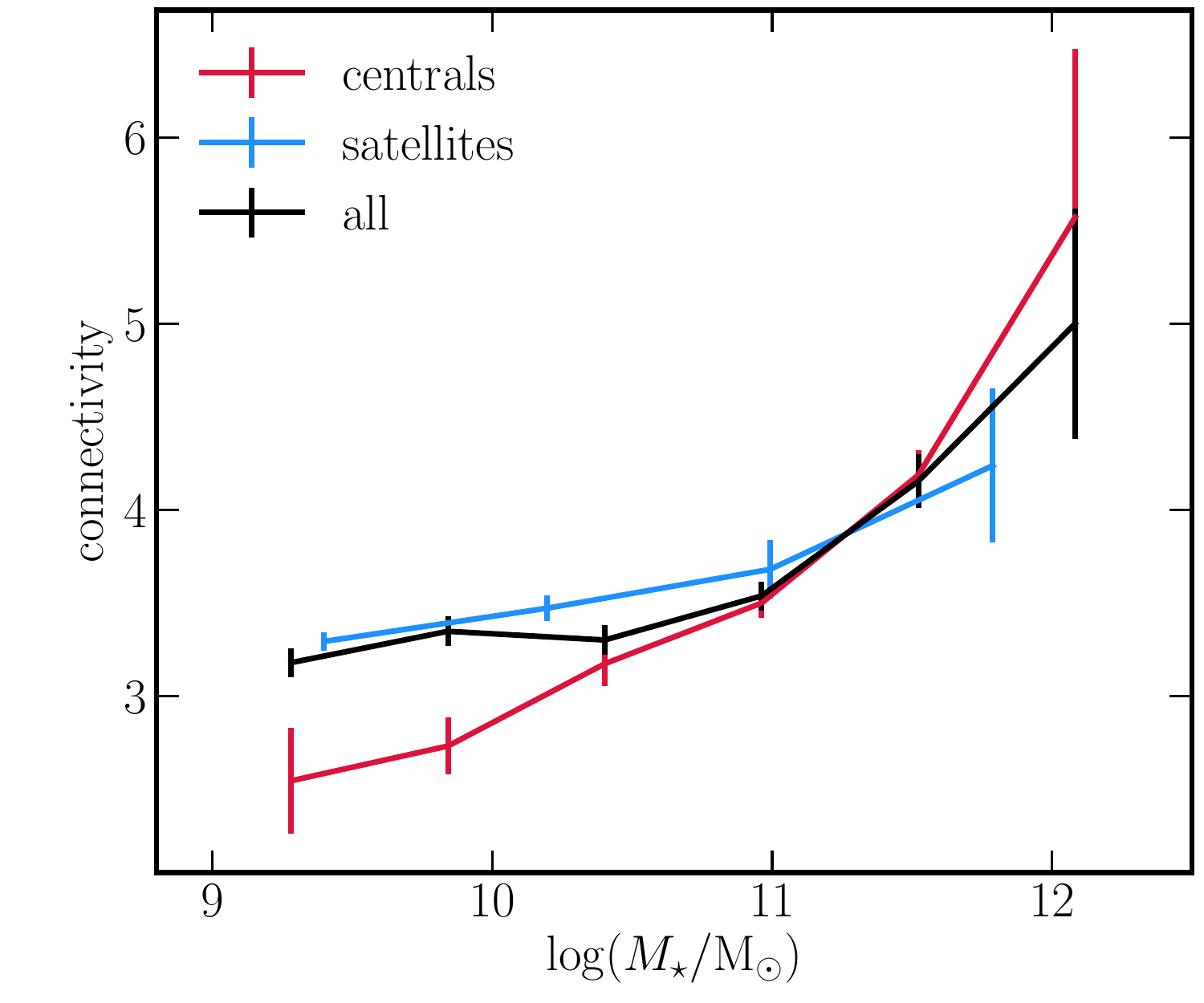}
\caption{Mean connectivity as a function of \mstar in \hagn at the persistence level $N_\sigma$ = 3
for centrals (in red), satellites (in blue) and the full population (in black).
The agreement with the expected trend (increasing connectivity with increasing \mstar) is stronger for the central. 
The elbow of the full population near $\mstar= 10^{11} \msun$ corresponds to the transition where it is dominated by satellites.
}
\label{fig:connect_mass_central}
\end{figure}

Let us briefly highlight the impact of the distribution of satellites on the connectivity--\mstar relation presented in the main text. The lower mass (satellite) galaxies dominate in number the population near the nodes of the cosmic web. 
This population has lower connectivity than the centrals, hence creating an elbow in the connectivity--\mstar relation for all galaxies, below which the connectivity shows only a weak dependence on \mstar.
This is seen on Figure~\ref{fig:connect_mass_central} which plots the connectivity as a function of \mstar for all galaxies (black line), centrals (red line) and satellites (blue line) in the \hagn simulation.
Globally, satellites do impact  figures like Figure~\ref{fig:connect_mass}, but for the sake of 
studying connectivity counts split by multiple physical parameters, we consider the full population in the main text.

\section{Multiplicity}
\label{sec:multiplicity}

Let us investigate the evolution of multiplicity with the physical parameters of connected galaxies. Multiplicity is complementary to connectivity, and is in principle a better proxy for the {\sl local} mass load as mentioned in Section~\ref{sec:cw}. On the other hand, multiplicity explicitly depends on the length over which the skeleton is smoothed. It is therefore less parameter free than the connectivity. The range of values it takes is also narrower, hence it provides less leverage over environment on larger scales (connectivity probes saddles points which can be some distance away from the nodes). It is also more complicated to predict from first principle (the geometry of accretion involves N-point correlations of extrema around peaks, as discussed in \citealt{codisetal2018}). 

\begin{figure}
\centering
\includegraphics[width=1.\columnwidth]{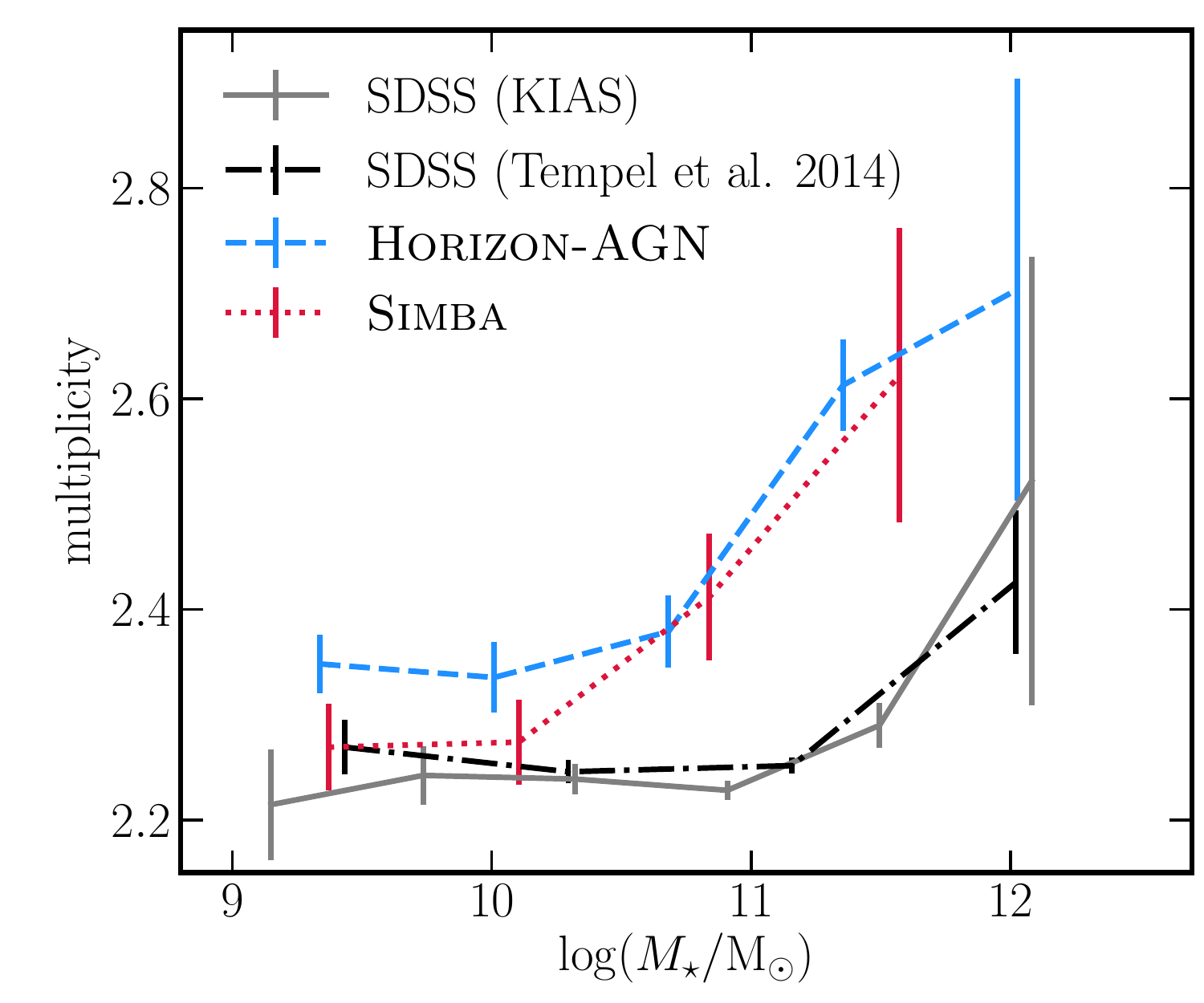}
\caption{Mean multiplicity as a function of \mstar in the SDSS (solid grey line), \hagn (dashed blue line) and \simba (dotted red line). In order to quantify our uncertainties, we also use the \protect\cite{Tempel2014} SDSS catalogue (dash-dotted black line). Both sets of simulations and post-processing of the raw SDSS data from both catalogues yield consistent measurements. Multiplicity increases with increasing \mstar in qualitative agreement between observations and simulations, in a qualitative agreement with the measured connectivity (see Figure~\ref{fig:connect_mass_all}).
}
\label{fig:multi_mass_all}
\end{figure}

Figure~\ref{fig:multi_mass_all} shows the multiplicity averaged over galaxies in bins of \mstar for the entire galaxy population in the two SDSS catalogues (solid grey and dashed-dotted black lines), \hagn (dashed blue line) and \simba (red dotted line). Multiplicity of galaxies is found to increase with increasing \mstar, in a qualitative agreement with the measured connectivity (see Figure~\ref{fig:connect_mass_all}).

\begin{figure}
\centering\includegraphics[width=\columnwidth]{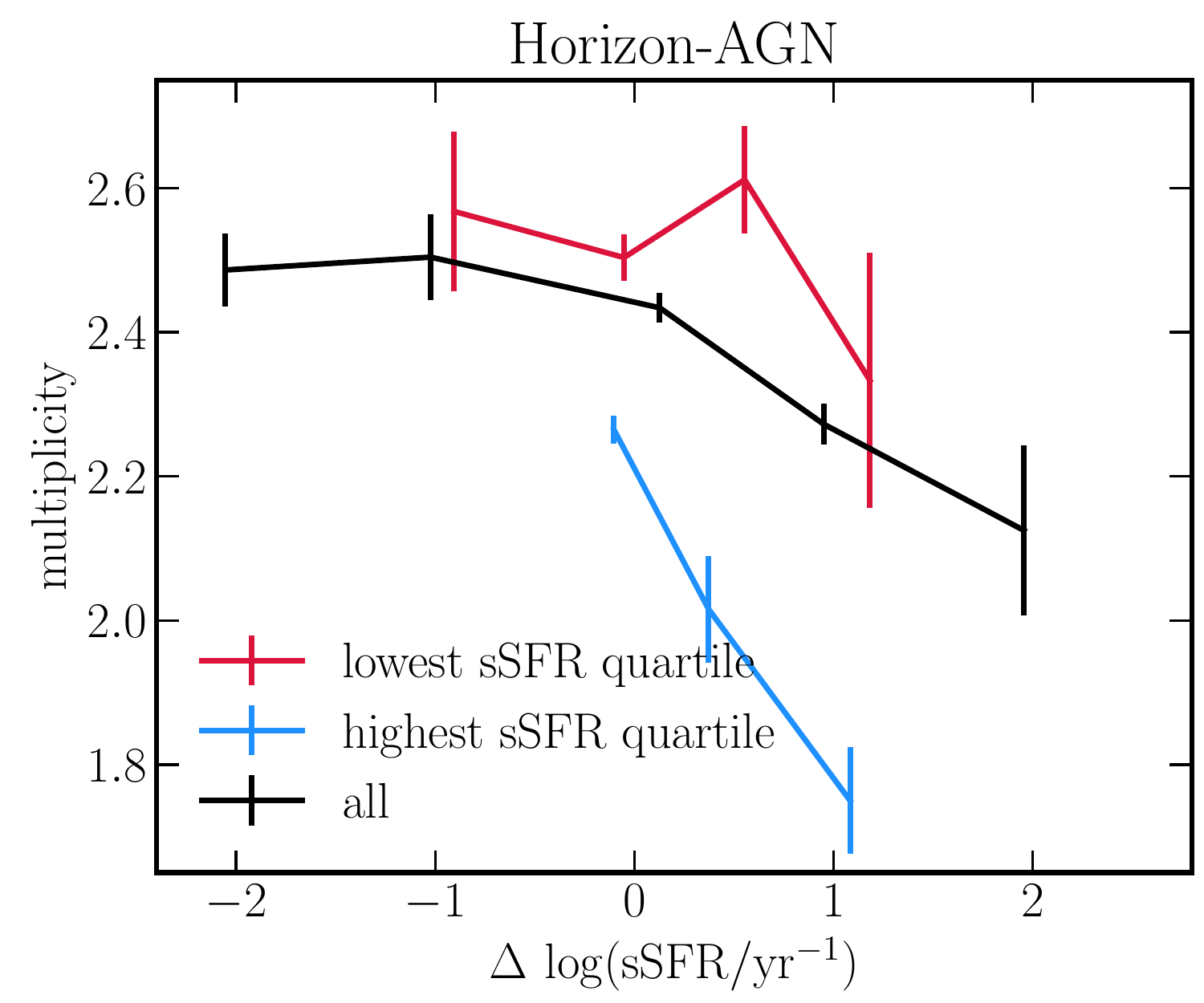}
\centering\includegraphics[width=\columnwidth]{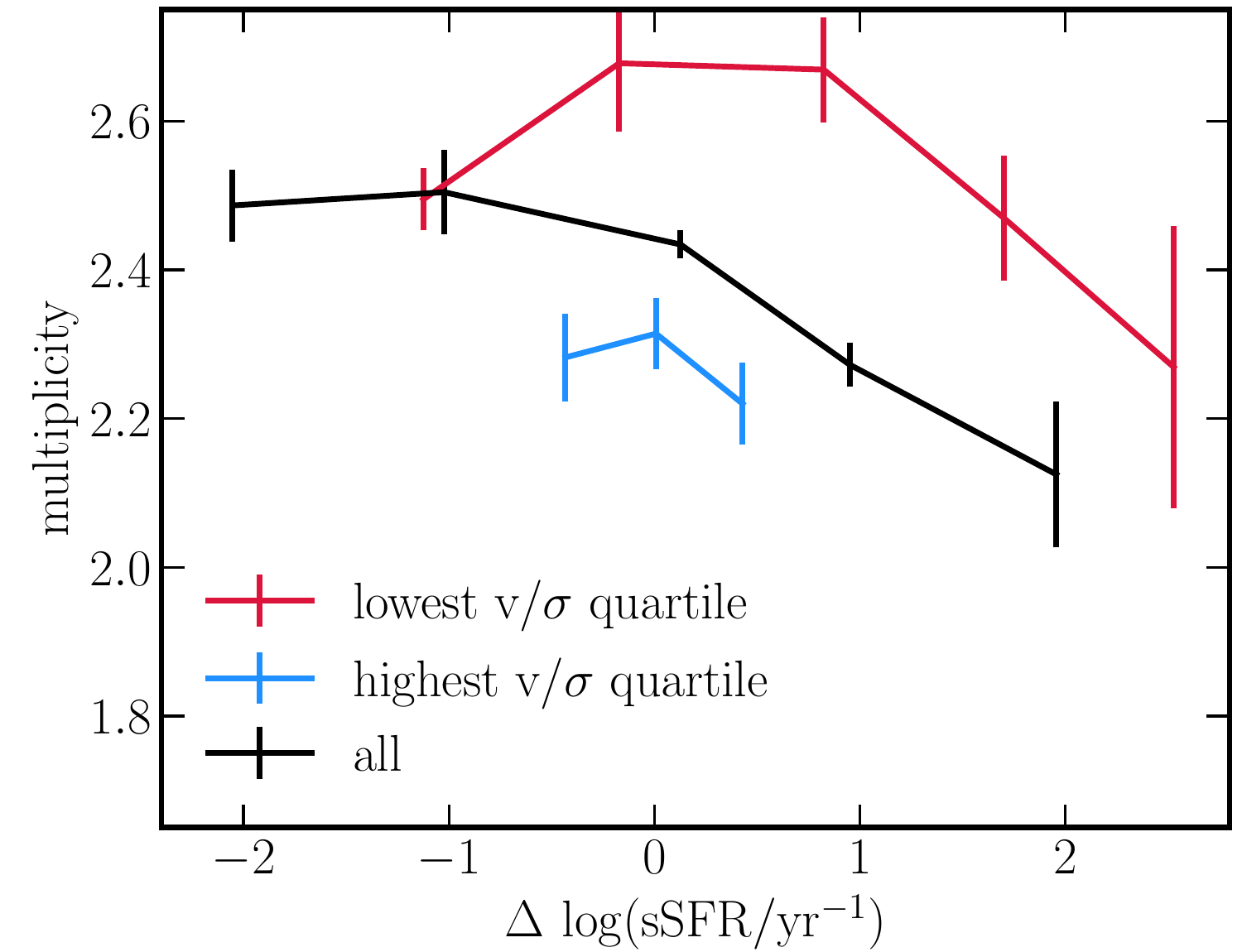}
\caption{Mean multiplicity as a function of the excess of \ssfr at given stellar mass in \hagn. 
The \textsl{top panel} is   split by star formation activity, while the \textsl{bottom panel}  by  morphology. The black solid line in each panel shows the mean multiplicity for all galaxies. 
Both passive (red line) and star-forming (blue line) galaxies with higher (lower) multiplicity tend to have lower (higher) \ssfr than the average at fixed \mstar of a given population. At fixed \ssfr excess, galaxies with low \ssfr tend to be more connected than star-forming ones.  
Similarly, both ellipticals (red line) and disk-dominated (blue line) galaxies with higher (lower) multiplicity tend to have lower (higher) \ssfr than the average at fixed \mstar of a given population. At fixed \ssfr excess, ellipticals are more connected than spiral galaxies.
Overall, galaxies with higher multiplicity  have lower \ssfr than the average population at  the same \mstar, regardless of their morphology or star formation activity.
}
\label{fig:multi_deltaSSFR_HzAGN}
\end{figure}

\begin{figure}
\centering\includegraphics[width=\columnwidth]{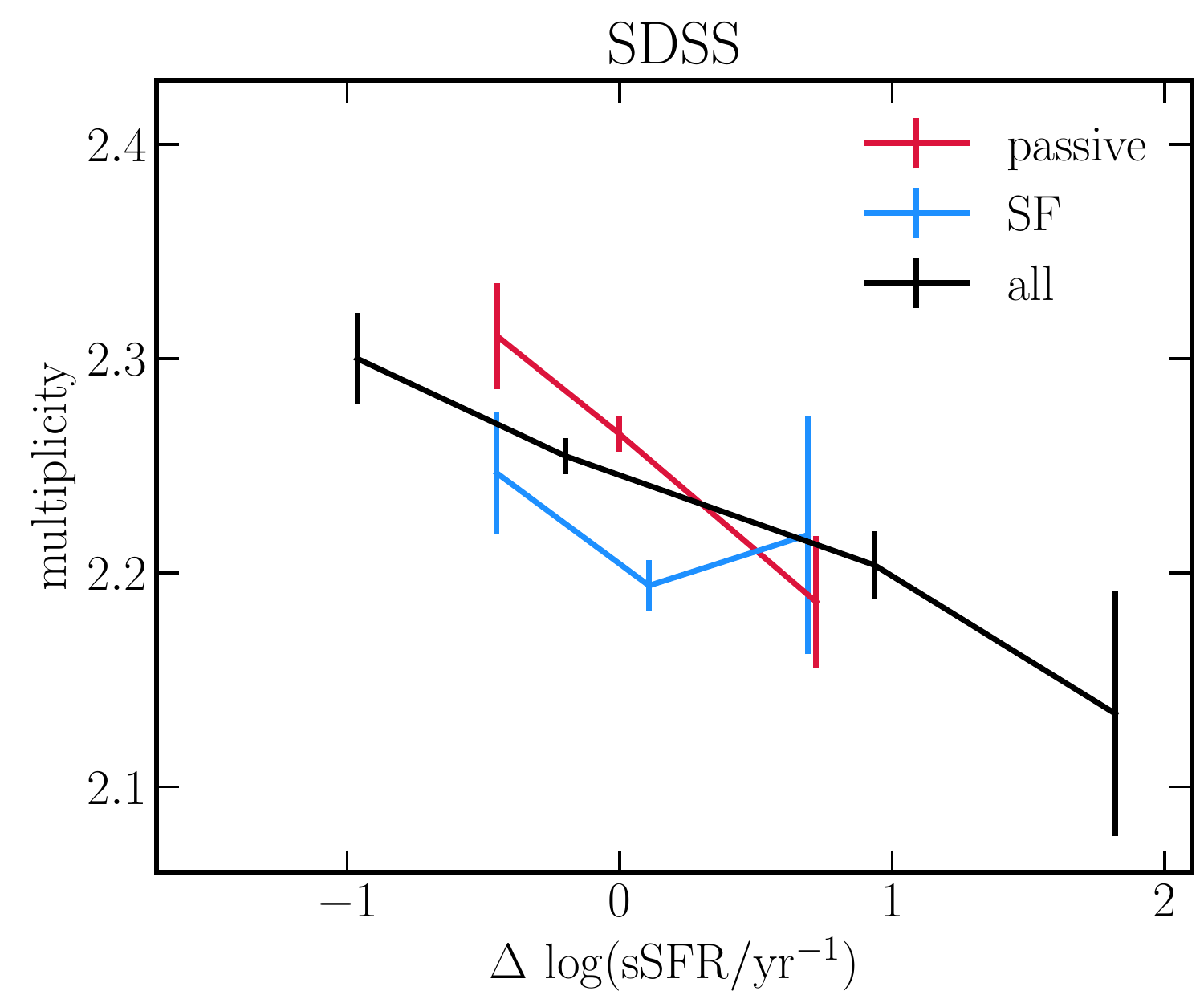}
\centering\includegraphics[width=\columnwidth]{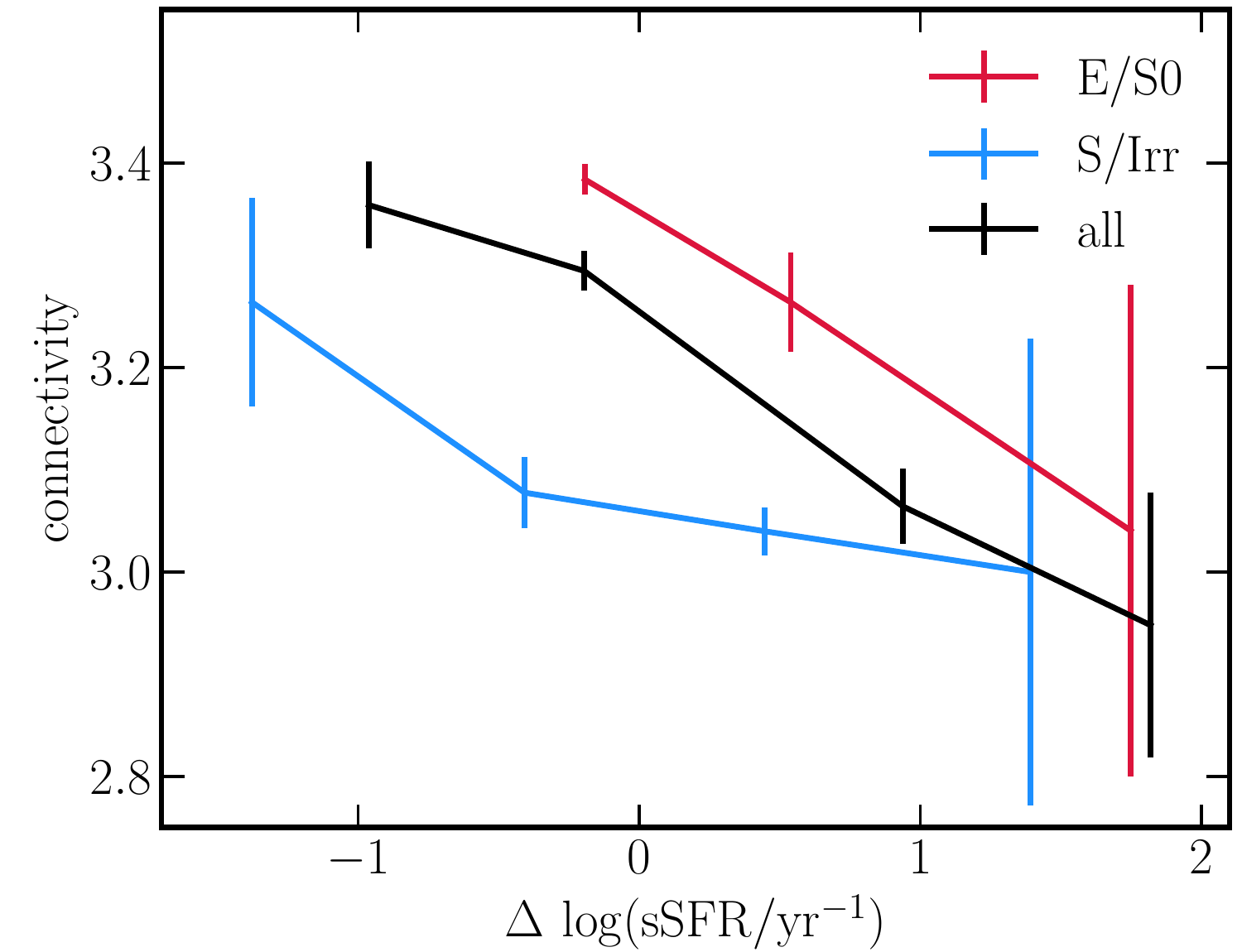}
\caption{Mean multiplicity as a function of the excess of \ssfr at given stellar mass for SDSS galaxies. 
The \textsl{top panel} is split by star formation activity of galaxies,  while the \textsl{bottom panel} by  morphology. The black solid line in each panel shows the mean multiplicity for all galaxies. 
As in figure \protect\ref{fig:multi_deltaSSFR_HzAGN}, both passive and star-forming galaxies with higher (lower) connectivity tend to have lower (higher) \ssfr than the average at fixed \mstar of a given population. At fixed \ssfr excess, passive galaxies tend to be more connected (locally) than star-forming ones.  
Similarly, both elliptical/S0 and spiral/irregular galaxies with higher (lower) multiplicity tend to have lower (higher) \ssfr than the average at fixed \mstar of a given population. At fixed \ssfr excess, elliptical/S0 galaxies tend to be more connected than galaxies with spiral/irregular morphology.  
}
\label{fig:multi_deltaSSFR}
\end{figure}

Figures~\ref{fig:multi_deltaSSFR_HzAGN} and \ref{fig:multi_deltaSSFR} show the multiplicity as a function of the \ssfr excess at given stellar mass, $\Delta \log ({\rm sSFR}/{\rm yr}^{-1})$, in \hagn and the SDSS, respectively. Qualitatively similar trends to those obtained for the connectivity are found when splitting by \ssfr and morphology (or \vsig in \hagn), for both simulations and for the observed data (see Figures~\ref{fig:connect_deltaSSFR} and \ref{fig:connect_deltaSSFR_HzAGN}).

\section{Connectivity in SIMBA}
\label{sec:simba_results}

\begin{figure}
\centering\includegraphics[width=\columnwidth]{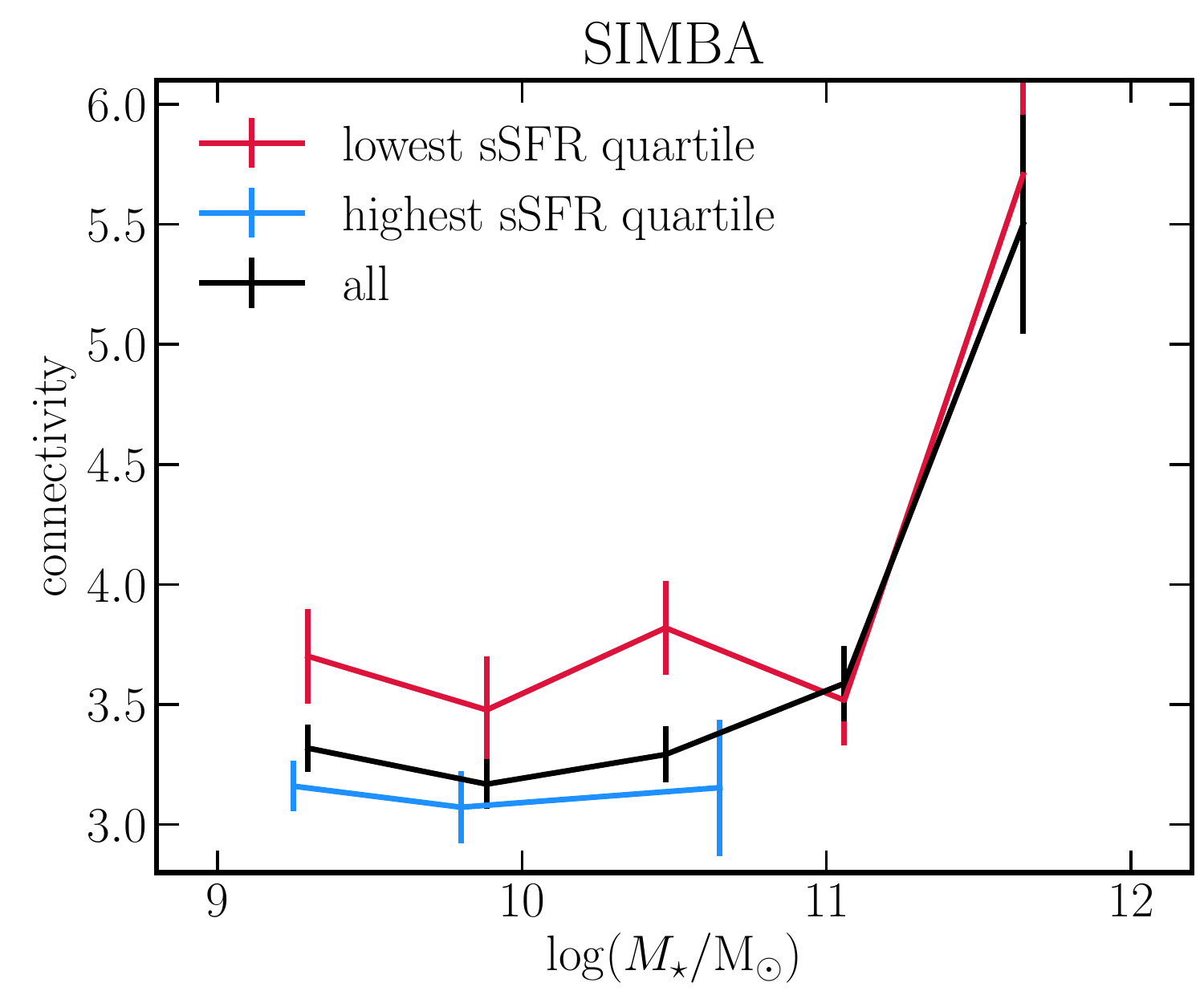}
\centering\includegraphics[width=\columnwidth]{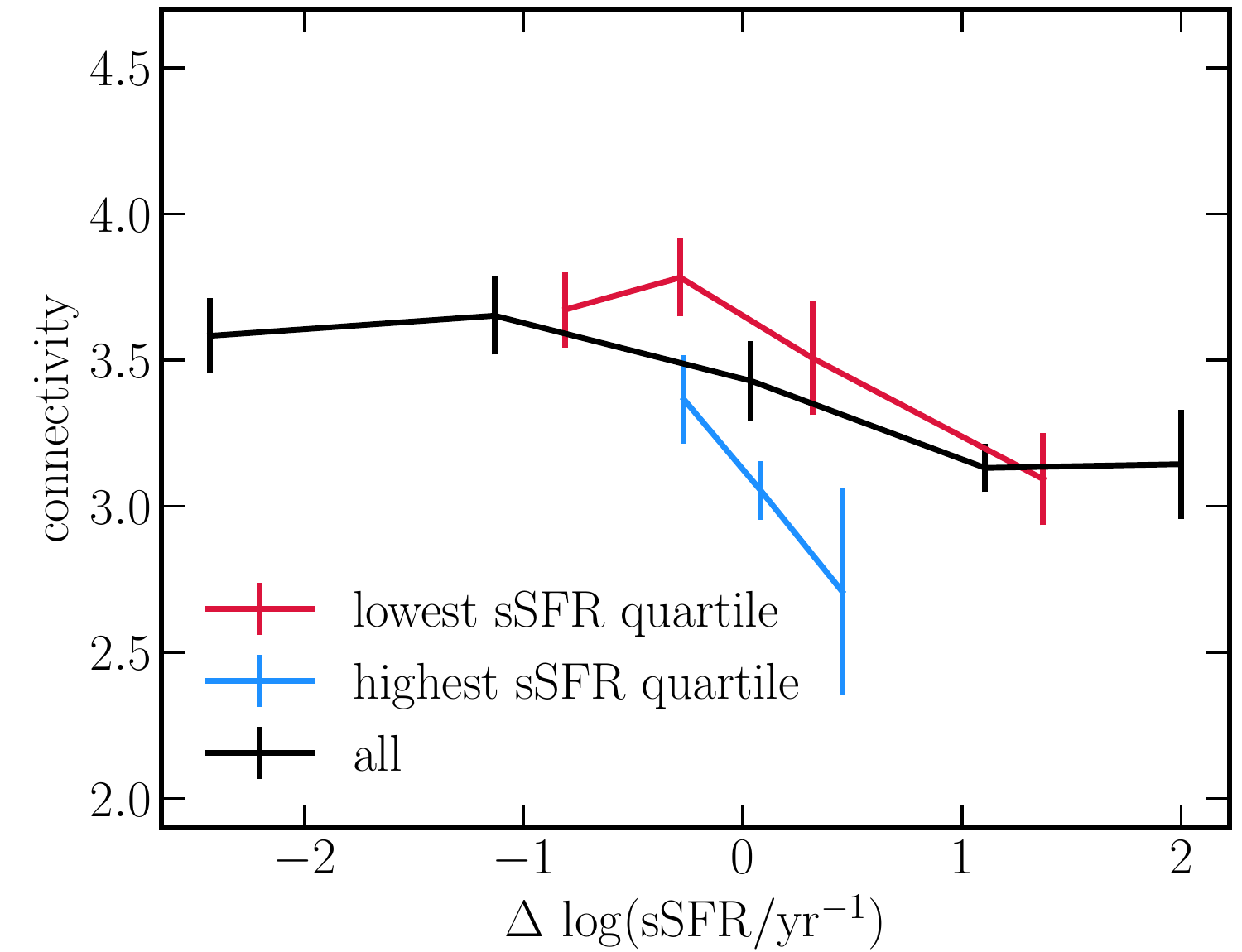}
\caption{Mean connectivity as a function of \mstar (\textsl{top panel}) and the \ssfr excess (\textsl{bottom panel}) for star-forming (blue lines) and passive (red lines) galaxies in \simba. The black solid line in each panel shows the mean connectivity for all galaxies at fixed mass.
Galaxies with low \ssfr have higher connectivity than star-forming galaxies (blue lines) at the same \mstar, in agreement with trends in the SDSS (see Figure~\ref{fig:connect_mass}) and \hagn (see Figure~\ref{fig:connect_mass_HzAGN}).
Both passive and star-forming galaxies with higher (lower) connectivity  have lower (higher) \ssfr than  average, at fixed \mstar of a given population. At fixed \ssfr excess, galaxies with low \ssfr are more connected than star-forming ones.  
Overall, galaxies with higher connectivity have lower \ssfr than the average population at the same \mstar, regardless of their star formation activity, in agreement with trends seen in observations (see Figure~\ref{fig:connect_deltaSSFR}) and \hagn (see Figure~\ref{fig:connect_deltaSSFR_HzAGN}).
}
\label{fig:connect_simba_ssfr}
\end{figure}

We finally present in Figure~\ref{fig:connect_simba_ssfr} the mean connectivity as a function of \mstar (top panel) and the \ssfr excess \ssfr at fixed \mstar (bottom panel) in the \simba simulation. 
Overall, the trends with \ssfr are qualitatively similar to those obtained in the SDSS and \hagn, albeit with larger uncertainties due to lower statistics. The variation of the mean connectivity with the excess  \ssfr is also flatter,  however, at fixed \ssfr excess, galaxies with low \ssfr continue to have higher connectivity than highly star-forming galaxies. The details of the baryonic physics modelling  likely impact  the measured trends. Assessing the origin of the differences  is beyond the scope of this paper.

\bsp	
\label{lastpage}
\end{document}